\documentclass[aip,jcp,reprint,showpacs,singlecolumn,superscriptaddress]{revtex4-1}

\usepackage{graphicx}
\usepackage{color}
\usepackage{amsmath}
\usepackage{comment}
\usepackage[export]{adjustbox}
\usepackage{placeins}
\newcommand{\be}{\begin{equation}}
\newcommand{\ee}{\end{equation}}
\newcommand{\bea}{\begin{eqnarray}}
\newcommand{\eea}{\end{eqnarray}}
\usepackage{dcolumn}
\usepackage{hyperref}
\usepackage{bm}
\usepackage{epsf}
\usepackage{subfigure}
\usepackage{epstopdf}%
\usepackage{amsfonts}
\usepackage{mattens} 

\begin{document}

\title{Effects of vibrational anharmonicity on molecular electronic conduction and thermoelectric efficiency}

\author{Hava Meira Friedman, Bijay Kumar Agarwalla and Dvira Segal}
\affiliation{Chemical Physics Theory Group, Department of Chemistry,
and Centre for Quantum Information and Quantum Control,
University of Toronto, 80 Saint George St., Toronto, Ontario, Canada M5S 3H6}

\date{\today}

\begin{abstract}
We study inelastic vibration-assisted charge transfer effects in two-site molecular junctions,
focusing on signatures of vibrational anharmonicity on the electrical characteristics
and the thermoelectric response of the junction.
We consider three types of oscillators: harmonic, anharmonic-Morse
allowing bond dissociation, and harmonic-quartic, mimicking a confinement potential.
Using a quantum master equation method which is perturbative in the electron-vibration
interaction we find that the (inelastic) electrical and thermal conductances can be largely affected
by the nature of the vibrational potential. In contrast, the Seebeck coefficient, the thermoelectric
figure-of-merit, and the thermoelectric efficiency beyond linear response,
conceal this information, showing a rather weak sensitivity to vibrational anharmonicity.
Our work illustrates that anharmonic (many-body) effects,
consequential to the current-voltage characteristics, are of little effect for the thermoelectric performance.
\end{abstract}

\maketitle


\section{Introduction}
\label{intro}

The interaction of electrons with nuclear degrees of freedom influences
the performance of molecular electronic junctions \cite{Scheer} by potentially supporting significant effects such as:
incoherent tunnelling processes, the development of hopping conduction \cite{Frisbie},
vibrational heating \cite{Natheat}, instability, and junction rupture \cite{lathaRup},
and the realization of intricate electron-electron and electron-vibration many-body phenomena \cite{TalK,NatStark}. 
Beyond electrical conductance, the Seebeck coefficient, which measures
the voltage that develops when a small temperature difference is applied, under the condition that the net charge current vanishes,
hands over information about the structure and energetics of molecular junctions.
It reveals, e.g., the nature of molecular orbitals hybridizing with the metal electrodes, and whether
the conductance is HOMO or LUMO dominated \cite{reddy,Reddy9,Reddy10,Reddy11,Tao,MalenE08, Latha13,reddy14}.

Theoretical descriptions of single-molecule electronic junctions essentially assume that molecular
vibrations are harmonic, as in the celebrated Anderson-Holstein (AH) model \cite{AH-Nitzan,WangAIP}, the
phonon-assisted donor-acceptor (DA) charge transfer model \cite{Lu,SiminePCCP, SimineINFPI,VonOppen,Bijay16},
or in multi-electronic state constructions \cite{ThossDA,Fabian3}.
The harmonic approximation is valid when atomic displacements are rather limited. It
allows one to solve the transport problem analytically---in certain limits---
and reach, e.g., the cumulant generating function, which provides closed expressions for the charge current
and high order cumulants, see e.g. Refs. \cite{AH-CGF1,AH-CGF2,BijayPRB15}.
It is important, however, to examine nanojunctions beyond the ideal harmonic-mode limit
and understand the role of vibrational anharmonicity on electronic transport through molecules.
Anharmonic effects are important when the applied bias voltage is high. Conducting electrons
then dispose significant amount of energy into the nuclear motion, resulting in large atomic displacements,
vibrational heating, and eventually bond dissociation.
As well, electrons in nanostructures may couple to
naturally-anharmonic degrees of freedom: molecular rotors, such as the torsional motion of two rings in the biphenyl molecule
\cite{cizek05,cizek11}, magnetic impurities \cite{Mn12,Ralph,MM11,MM12}, molecular conformations
\cite{troisi03,LathaConfor}.

\begin{figure}
\hspace{16mm}
\includegraphics[width=13cm, angle=-90]{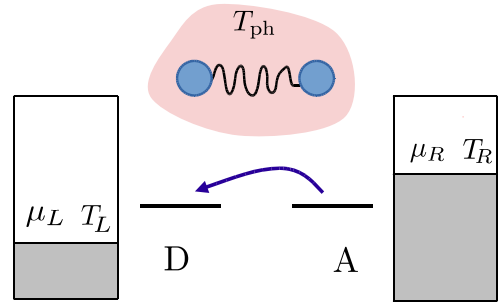}
\vspace{-80mm}
\caption{Scheme of a voltage-biased donor-acceptor molecular junction.
Electron hopping between the D and A sites
is coupled to a specific (primary) molecular vibration, modeled by an harmonic or an anharmonic oscillator.
The primary oscillator may dissipate its energy to a secondary phononic (harmonic) environment of temperature $T_{ph}$,
represented by the shaded region.
}
\label{Fig1}
\end{figure}


So far, the investigation into the role of anharmonic oscillations in electron transport in molecules has received little attention.
It was demonstrated in Ref. \cite{Koch1} that in the sequential-tunneling regime
steps in the I-V (current-voltage) characteristics, the result of (harmonic) vibrational excitations, 
split into a multitude of steps under the Morse potential. 
Other unique signatures of vibrational anharmonicity, as revealed in Ref. \cite{Koch1}, were bias-dependent
broadening of vibrational features in conductance and the development of negative differential conductance.
Current-induced molecular dissociation rates were calculated in Ref. \cite{Koch2}.
I-V characteristics with effective anharmonic (double-well) vibrational potentials were examined in Refs. \cite{Grifoni06, Brandes07}
showing rich effects. In Ref. \cite{Peskin05}, the degree of anharmonicity
was demonstrated to affect the rate of electron tunneling in donor-bridge-acceptor complexes.
Nevertheless, unlike the harmonic case, analytical results for transport behavior in anharmonic junctions are
missing, given the complexity of the problem.

Motivated to examine effects of vibrational anharmonicity on electron transport characteristics
in an analytically tractable model, we had recently introduced the so-called spin-fermion model \cite{SiminePCCP}.
In this setup, electrons in the junction couple to a highly anharmonic impurity mode, which consists of only two
states, replacing the full harmonic manifold.
Based on this model,
we had examined the role of mode harmonicity/anharmonicity
on vibrational heating, cooling, and instability, under high voltage biases \cite{SiminePCCP,SimineINFPI}, then
analyzed the impact of mode anharmonicity on current blockade physics  \cite{SimineAH}.
Moreover, in Refs. \cite{SiminePCCP,BijayPRB15} we derived the cumulant generating function of
the phonon-assisted donor-acceptor model with either a harmonic mode or a two-state impurity.
We then showed that while the inelastic current and its cumulants
exhibited significant signatures of molecular anharmonicity, the thermoelectric energy
conversion efficiency was indifferent to the nature of the mode; it was precisely identical
when working with either a harmonic local mode, or a two-level system \cite{BijayPRB15,BijayDABeil}.
This result was obtained under the weak electron-vibration coupling approximation, 
but allowing for strong metal-molecule hybridization.

This remarkable result, namely, the exact correspondence
of the thermoelectric performance in DA junctions
with either harmonic or two-state modes, calls for additional investigations.
Naturally, one questions whether this agreement is a consequence of the fact that a two-state impurity is
characterized by (obviously) a single energy gap, similarly to the harmonic mode in the weak coupling limit,
when multi-quanta processes are disallowed. Alternatively, this indifference to the nature
of the vibrational potential may not be
coincidental, rather reflecting that measures related to ratios of charge and energy currents only weakly depend
on the anharmonic potential.

The objective of the present study is to examine the effects of realistic anharmonic vibrational potentials
on inelastic conduction within the phonon-assisted donor-acceptor model of Fig. \ref{Fig1}, by investigating the model's
I-V characteristics and thermoelectric behavior.
In this construction, electron transfer between the D and A sites is assisted by a particular (primary) vibrational mode,
isolated, or coupled to a secondary phonon bath. The primary vibrational oscillator
may be made anharmonic, and we consider here three representative potentials: harmonic, anharmonic-Morse
where symmetry between mode compression and mode stretching is broken,
and harmonic-quartic potential, describing less flexible bonds (relative to the harmonic case).
Considering these three types of DA junctions,
we study the system's linear response transport coefficients,
high-bias I-V characteristics, 
and nonlinear thermoelectric efficiency, to identify the role of vibrational anharmonicity on inelastic transport.

We explore transport characteristics of our model using
a quantum master equation (QME) method, perturbative in
the electron-vibration coupling but exact to all order in the metal-molecule hybridization \cite{SiminePCCP}.
Remarkably, we find that in agreement with previous results on the two-state anharmonic mode \cite{BijayPRB15,BijayDABeil},
while the I-V characteristics significantly deviate under different anharmonic potentials,
the Seebeck coefficient and the thermoelectric efficiency, even beyond linear response, manifest a weak sensitivity
to the vibrational potential.

The paper is organized as follows. We introduce the model in Sec. \ref{model}.
In Sec. \ref{method}, we apply a master equation method to the molecular electronic junction problem
and explain how we calculate transport properties.
Simulation results are presented in Sec. \ref{result}. We summarize our work in Sec. \ref{Summary}.
Throughout the paper we work with units where $\hbar=1$, $k_B=1$ and $e=1$.

\section{Model}
\label{model}

We consider a prototype molecule with two electronic states, denoted by donor (D) and acceptor (A)
following chemistry literature, see Fig. \ref{Fig1}.
The molecule bridges two metal electrodes comprising non-interacting electrons.
Electron transfer between D and A takes place by an inelastic process, with electrons exchanging energy
with the primary molecular oscillator, which is itself coupled to a secondary phonon bath.

We employ below a quantum kinetic master equation approach which can be rigorously derived from the Liouville equation
under the assumptions of weak system-bath coupling, Markovian environments, and secular dynamics \cite{BookOQS}.
Projection operator approaches are developed based on the conceptual separation of the Hamiltonian into a subsystem plus bath,
\bea
\hat H=\hat H_S+\hat H_B + \hat V.
\eea
In this work, the particular primary oscillator serves as the subsystem. The environment
$\hat H_B$ comprises two baths:
a fermionic bath consisting of the electronic degrees of freedom (molecular states plus metals),
and a bosonic bath collecting the secondary phonon modes.
In the energy basis, the subsystem Hamiltonian and the interaction with the environment are written as
\bea
\hat H_S&=&\sum_{n}E_n|n\rangle \langle n|,
\nonumber\\
\hat V&=& \hat B\otimes \hat S = \hat B\otimes \sum_{m,n} S_{m,n}|m\rangle\langle n|.
\eea
$S_{m,n}\equiv\langle m|\hat S| n \rangle$ with $\hat S$ a subsystem operator.
$\hat B$ is an operator of the baths including two contributions,
$\hat B=\hat B_{el}+\hat B_{ph}$.
In Section \ref{modelsubs}, we specify the subsystem- the molecular oscillator.
In Section  \ref{modelbaths}, we describe the electronic and bosonic thermal baths.


\subsection{Subsystem: primary oscillator}
\label{modelsubs}

The single molecular oscillator, representing molecular nuclear motion, defines our subsystem.
Using mass-weighted coordinates, displacement $\hat x$ and momentum $\hat p$,
the corresponding Hamiltonian is written as
\bea
\hat H_S=\frac{\hat p^2}{2} + U(\hat x), 
\eea
with $U(\hat x)$ the potential energy function, not necessarily harmonic.
Unless otherwise specified, we assume that the (dimensionless)
subsystem's interaction operator $\hat S$ takes the form
\bea
\hat S=  \sqrt{2\omega_0}\hat x= \hat b_0^{\dagger}+\hat b_0.
\eea
Here $\omega_0$ is a characteristic frequency of the subsystem, $\hat b_0^{\dagger}$ ($\hat b_0$)
are creation (annihilation) bosonic operators.
We consider three models for the primary oscillator: harmonic, Morse, and harmonic-quartic.

\noindent {\bf 1. Harmonic oscillator}.
The Hamiltonian $\hat H_S=\frac{\hat p^2}{2} + \frac{1}{2}\omega_0^2 \hat x^2$
supports the eigenenergies and matrix elements
\bea
E_n&=&\left(n+\frac{1}{2}\right)\omega_0, \,\,\, n=0,1,2,..
\nonumber\\
|S_{m,n}|^2&=&(n+1)\delta_{m,n+1}+ n\delta_{m,n-1}.
\label{eq:HO}
\eea

\noindent {\bf 2. Morse oscillator}.
This potential is defined in terms of the dissociation energy $D$ and a width parameter $\alpha$,
$\hat H_S=\frac{\hat p^2}{2} + D(e^{-\alpha \hat x}-1)^2$.
At small displacements, the potential can be approximated by a harmonic model of
frequency $\omega_0=\alpha\sqrt{2D}$.
The eigenenergies of the model and the matrix elements of $\hat S$ take a closed form, 
\bea
&&E_n=\omega_0\left(n+\frac{1}{2}\right)  -\frac{\omega_0^2}{4D}\left(n+\frac{1}{2}\right)^2, \,\,\,\,\, n=0,1,2,.., n_{max}.
\nonumber\\
&&| S_{m,n}|^2=2\lambda\left(\frac{2 (-1)^{m-n+1}}{(m-n)(2\tilde\lambda-n-m)}\right)^2
\nonumber\\
&&\times
\frac{(\tilde \lambda-n)(\tilde \lambda -m)\Gamma(2\tilde \lambda-m+1)m!}{\Gamma(2\tilde \lambda-n+1)n!} \,\,\,\,\  (m>n),
\label{eq:Morse}
\eea
with $\lambda=2D/\omega_0$ and $\tilde\lambda=\lambda-1/2$.
The Morse potential breaks the symmetry between mode stretching and compression,
as reflected by the full matrix $\hat S$.

\noindent {\bf 3. Harmonic-Quartic (HQ) oscillator}.
We introduce a quartic contribution on top of the harmonic potential function,
$\hat H_S=\frac{\hat p^2}{2} + \frac{1}{2}\omega_0^2 \hat x^2 + a_4 \omega_0^4 \hat x^4$.
Here, $a_4$ is the anharmonic coefficient, with physical dimension of inverse energy.
The HQ potential describes symmetric inflexible-confined motion; $x^{2d}$ approximates a 1D box for large (positive integer) $d$.
Below we use a DVR algorithm \cite{DVR} to receive $E_n$ and the matrix elements $ S_{m,n}$ of the HQ model.

\begin{figure*}
\hspace{4mm}
\includegraphics[width=15cm]{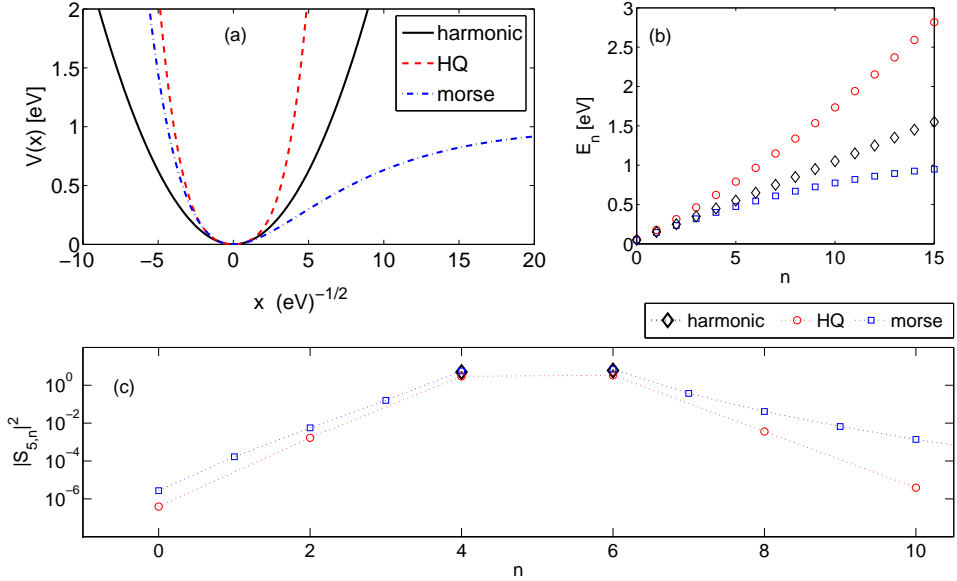}
\caption{
(a) Potential energy for the harmonic, HQ and Morse oscillators as a function of the mass-weighted coordinate $x$.
(b) Eigenenergies of the three oscillators.
(c) An example of coupling matrix elements $| S_{5,n}|^2$.
We used $\omega_0=0.1$ eV, dissociation energy (Morse) $D=1$ eV, and $a_4=1$ 1/eV (HQ potential).
}
\label{Fig2}
\end{figure*}

In Fig. \ref{Fig2} we depict the three vibrational potentials,
the corresponding eigenenergies, and examples for matrix elements of $\hat S$.
The fundamental distinctions between the two types of anharmonicity are:
(i) The Morse (HQ) potential supports energy levels with energy spacings smaller (larger) than the harmonic limit $\omega_0$,
see panel b.
(ii) The HQ potential maintains an even symmetry around the equilibrium position, similarly to the harmonic model.
As a result, the eigenfunctions of the HQ potential acquire a definite (even, odd) symmetry, thus matrix elements of
$\hat S$ survive only between states of opposite symmetry.
In contrast, the Morse potential is missing a definite symmetry, thus it allows transitions between any pair of states, see panel c.

\subsection{Reservoirs: electronic and phononic baths}
\label{modelbaths}

The primary molecular oscillator, defined as $\hat H_S$, couples
to electronic (el) degrees of freedom and to secondary-harmonic modes---a phononic (ph) environment,
\bea
\hat H_B = \hat H_{el} + \hat H_{ph}, \,\,\,\, \hat V=\hat V_{el}+ \hat V_{ph}.
\eea
We recall that $\hat V=\hat S\otimes \hat B$,
$\hat B=\hat B_{el}+\hat B_{ph}$. 
The phononic environment includes independent harmonic modes, bilinearly coupled to the primary oscillator,
\bea
\hat H_{ph}&=&\sum_{k}\omega_k \hat b_k^{\dagger} \hat b_k,
\nonumber\\
\hat
V_{ph}&=& \left(\hat b_0^{\dagger}+\hat b_0 \right)\sum_k \nu_k \left(
\hat b_k^{\dagger}+\hat b_k\right),
\eea
$\hat b_k^{\dagger}$ ($\hat b_k$) as bosonic creation (annihilation) operators for the $k$th mode
of frequency $\omega_k$.
The electronic reservoir includes both metals and the molecular electronic states
\bea
\hat H_{el}
&=& {\epsilon}_d \hat c_d^{\dagger} \hat c_d + {\epsilon}_a \hat c_a^{\dagger} \hat c_a
+ \sum_{l\in L} {\epsilon}_l \hat c_l^{\dagger} \hat c_l
+ \sum_{r\in R} {\epsilon}_r \hat c_r^{\dagger} \hat c_r
\nonumber\\
&+& \sum_{l \in L } v_{l} (\hat c_l^{\dagger} \hat c_d \!+\! \hat c_d^{\dagger} \hat c_l) \!+\! \sum_{r \in R} v_{r} (\hat c_r^{\dagger} \hat c_a \!+\! \hat c_{a}^{\dagger} \hat c_r).
\label{eq:Hel}
\eea
Here, $\epsilon_{d}, \epsilon_{a}$ are the donor and acceptor site energies, coupled to
the left $L$ and right $R$ metal leads by real-valued hopping elements $v_l$ and $v_r$, respectively.
$\hat c^{\dagger}$ and $\hat c$ are fermionic creation and annihilation operators.
The interaction between electrons in the junction and the primary vibrational mode is given by the ``off-diagonal" model,
\be
\hat V_{el}= g [\hat c_{d}^{\dagger} \hat c_{a} + \hat c_{a}^{\dagger} \hat c_d ]  (\hat b_0^{\dagger} + \hat b_0).
\label{eq:Hevib}
\ee
Note that we do not include here a direct-elastic electronic tunneling term between the D and A states.
This contribution can be accommodated approximately-separately, as a Landauer term to the current, see Appendix B.

The electronic Hamiltonian (\ref{eq:Hel}) can be diagonalized
and expressed in terms of new fermionic operators, $\hat a_l$ and $\hat a_r$.
In the new basis Eqs. (\ref{eq:Hel})-(\ref{eq:Hevib}) are given by
\bea
\hat H_{el}&=& \sum_{l}\epsilon_l \hat a_l^{\dagger}\hat a_l + \sum_r \epsilon_r \hat a_r^{\dagger}\hat a_r.
\nonumber\\
\hat V_{el}&=& g\sum_{l,r}\left[ \gamma_l^{*}\gamma_r \hat a_l^{\dagger}\hat a_r
+ \gamma_r^* \gamma_l \hat a_r^{\dagger}\hat a_l \right]
(\hat b_0^{\dagger}+\hat b_0),
\label{eq:VelD}
\eea
allowing us to identify the electronic operators,
\bea
\hat B_L = g\sum_{l,r} \gamma_l^{*}\gamma_r \hat a_l^{\dagger}\hat a_r, \,\,\,\,\,\,
\hat B_R = g\sum_{l,r} \gamma_r^{*}\gamma_l \hat a_r^{\dagger}\hat a_l,
\label{eq:BLBR}
\eea
responsible for electron hopping from the right compartment to the left one, and its
hermitian conjugate $\hat B_R=\hat B_L^{\dagger}$, transferring electrons from the left terminal to the right side.
The coefficients, e.g., for the $L$ set, are
\bea
\gamma_l=\frac{v_l}{\epsilon_l-\epsilon_d + i\Gamma_L(\epsilon)/2}.
\label{eq:diag}
 \eea
Note that we ignore the real-principal value term---responsible for a small energy shift of $\epsilon_{d,a}$. Here,
$\Gamma_{\nu}(\epsilon)=2\pi \sum_{j\in \nu} v_j^2\delta(\epsilon-\epsilon_j)$.
In what follows, we take this hybridization as a constant independent of energy, consistent with the
omission of the real part of the self energy.
The expectation values of the exact eigenstates, with respect to the electronic density matrix satisfy
\bea \langle \hat a_j^{\dagger}\hat a_{j'}\rangle=\delta_{j,j'}f_\nu(\epsilon_j),\,\,\,\,\,\,\,\ j\in \nu
\eea
with $f_{\nu}(\epsilon)=[\exp(\beta_{\nu}(\epsilon-\mu_{\nu}))+1]^{-1}$ as
the Fermi distribution function at inverse temperature $\beta_{\nu}=T_{\nu}^{-1}$ and chemical potential $\mu_{\nu}$, $\nu=L,R$.
Eq. (\ref{eq:VelD}) indicates that the following spectral density functions determine the subsystem's (oscillator) dynamics,
\bea J_{\nu}(\epsilon)&=&2\pi
g\sum_{j\in\nu}|\gamma_{j}|^2\delta(\epsilon_{j}-\epsilon).
\label{eq:JLR}
\eea
Using Eq. (\ref{eq:diag}), it can be shown that
the spectral functions take a Lorentzian lineshape centered about $\epsilon_{d,a}$,
\bea
J_L(\epsilon)&=&g\frac{\Gamma_L}{(\epsilon-\epsilon_d)^2+\Gamma_L^2/4}
\nonumber\\
J_R(\epsilon)&=&
g\frac{\Gamma_R}{(\epsilon-\epsilon_a)^2+\Gamma_R^2/4}.
\label{eq:spec}
 \eea
%
Below we show that these functions are the central building block in the transition rate constants between vibrational states,
constructing the expressions for electrical and energy currents.

For later use, we also separate the electronic Hamiltonian into the $L$ and $R$ compartments,
$\hat H_{\nu}= \sum_{j\in \nu} \epsilon_j \hat a_j^{\dagger}\hat a_{j}$,
and define the number operators $\hat N_{\nu}=\sum_{j\in \nu} \hat a_j^{\dagger}\hat a_{j}$.


\section{Method}
The purpose of this section is to outline a unified formalism
for the calculation of both the subsystem (vibration) dynamics
and the electron transport characteristics (currents), far from equilibrium.
In Sec. \ref{A}, we review the principles of a standard projection operator approach that hands over
equations of motion for the reduced density matrix.
In Sec. \ref{B}, we clarify that the characteristic function for transport can be evaluated in an analogous manner,
by writing it down as a trace over a counting-field dependent reduced density matrix.

\label{method}

\subsection{Population Dynamics: vibrational mode}
\label{A}

The molecular oscillator is identified as the subsystem, and it is interacting with electronic and phononic baths.
The reduced density matrix of the oscillator can be obtained from projection operator approaches by
making standard approximations:
weak subsystem-bath coupling, Markovianity of the electronic and phononic environments, secular approximation
for decoupling population and coherence dynamics, and
working with models satisfying $\langle \hat B_{el/ph}\rangle =0$.
Under these approximations,  the population $p_n$ of the (subsystem) state $n$ obeys a quantum kinetic equation \cite{BookOQS}
\bea
\dot p_n(t) = -p_n(t) \sum_m k_{n\to m} + \sum_m k_{m\to n} p_m(t),
\label{eq:pop}
\eea
with rate constants
\bea
k_{n \to m}=
 | S_{m,n}|^2 \int_{-\infty}^{\infty} d\tau e^{i (E_n-E_m) \tau}  \langle \hat B(\tau)\, \hat B(0) \rangle.
\label{eq:kF}
\eea
%
The $\hat B$ operators are written in the interaction representation,
$\hat B(\tau)=e^{i\hat H_0\tau}\hat Be^{-i\hat H_0\tau}$ with $\hat H_0=\hat H_S+\hat H_B$.
Averages are calculated with respect to the initial state of the baths $\hat \rho_B=\hat \rho_{el}\otimes\hat \rho_{ph}$,
$\langle \hat A(t)\rangle\equiv {\rm Tr}_B[\hat A(t) \hat \rho_B]$
$\hat \rho_{el}=\hat \rho_L\hat \rho_R$ with
$\hat \rho_{\nu}=e^{-\beta_{\nu}(\hat H_{\nu}-\mu_{\nu}\hat N_{\nu})}/
{\rm Tr}_{\nu}[ e^{-\beta_{\nu}(\hat H_{\nu}-\mu_{\nu}\hat N_{\nu})}]$,
see definitions at the end of Sec. \ref{modelbaths}.

Since $\hat B=\hat B_{el}+\hat B_{ph}$, 
and from Eq. (\ref{eq:BLBR}) $\hat B_{el}=\hat B_{L}+\hat B_R$, the rate constants are additive in the different processes,
\bea
k_{n\to m}= k_{n\to m}^{L \to R} + k_{n\to m}^{R \to L} + k_{n\to m}^{ph}.
\label{eq:ratessum}
\eea
The electronic rates  ($\nu=L,R$, $\bar\nu=R,L$) are
\bea
k_{n \to m}^{\nu\to \bar \nu}=
 | S_{m,n}|^2 \int_{-\infty}^{\infty} d\tau e^{i (E_n-E_m) \tau}  \langle \hat B_{\nu}(\tau)\, \hat B_{\bar \nu}(0) \rangle
\nonumber\\
\label{eq:kFe}
\eea
with \cite{SiminePCCP}
\begin{widetext}
\bea
k_{n \to m}^{L \to R} &=& |S_{m,n}|^2\int_{-\infty}^{\infty}\!
\frac{d\epsilon}{2\pi} f_L(\epsilon)
[1\!-\!f_R(\epsilon\!+E_{nm})] J_L(\epsilon) J_R(\epsilon\!+\!E_{nm}), \nonumber \\
k_{n\to m}^{R \to L} &=& | S_{m,n}|^2 \int_{-\infty}^{\infty}
\!\frac{d\epsilon}{2\pi} f_R(\epsilon) [1\!-\!f_L(\epsilon\!+\!E_{nm})] J_R(\epsilon)
J_L(\epsilon\!+\!E_{nm}).
\label{eq:kel}
\eea
\end{widetext}
Here, $E_{nm}\equiv E_n-E_m$, where as we recall, $E_n$ are the eigenenergies of the primary oscillator. 
The electronic rate constants are given in terms of the Fermi-Dirac functions
and the spectral density functions of the left and right electronic leads
(involving the molecular electronic states).
These terms are nonzero when (i)  both leads are not fully occupied or empty,
 and (ii) the overlap between the spectral functions, differing by one quanta of energy, is non-negligible.
Because of the assumed weak electron-phonon coupling, each electron tunnelling process involves
absorption/emission of a single vibrational quanta.

The phonon bath-induced rates are evaluated with the average taken over the canonical distribution
$\hat \rho_{ph}=e^{-\beta_{ph}\hat H_{ph}}/Z_{ph}$ with the partition function
$Z_{ph}={\rm Tr}[e^{-\beta_{ph}\hat H_{ph}}]$ and
the inverse temperature $\beta_{ph}=1/T_{ph}$,
\bea
k_{n\to m}^{ph}=\Gamma_{ph}(|E_{mn}|)n_{ph}(E_{mn}) sgn(E_{mn}). 
\label{eq:kph}
\eea
The vibration-phonon bath coupling energy is
\bea
\Gamma_{ph}(\omega)=2\pi\sum_{k}\nu_k^2\delta(\omega-\omega_k),
\eea
later taken as an energy-independent constant \cite{commGaph}.
$n_{ph}(\omega)=\left[e^{\beta_{ph}\omega}-1\right]^{-1}$
is the Bose-Einstein occupation factor.
We work with $\Gamma_{ph}$ large enough so as to satisfy $k_{n\to m}<k_{m\to n}$ for $m>n$
and rule out the phenomenon of vibrational instability \cite{SiminePCCP},
the uncontrolled bias-induce heating of the vibration.

For later use, it is convenient to organize  the population dynamics (\ref{eq:pop})
in a matrix form
\bea
|\dot p\rangle &=&{\cal L} |p\rangle
\nonumber\\
&=&
\left({\cal L}_{L \to R} + {\cal L}_{R\to L} + {\cal L}_{ph}\right)
|p\rangle,
\label{eq:dyn}
\eea
with $|p\rangle$ a vector collecting the subsystem population,  ${\cal L}$ is the so-called Liouvillian.

It is useful to recall that if the primary mode is harmonic, only transitions between neighboring states
survive according to Eq. (\ref{eq:HO}).
The population dynamics then simplifies to
\bea
\dot p_n &=& -\left[nk_{n\to n-1 } + (n+1)k_{n\to n+1}\right] \,p_n
\nonumber\\
&+& (n+1)k_{n+1\to n}\,p_{n+1} +(n-1)k_{n-1\to n}\,p_{n-1}.
\eea
In contrast, the Morse and HQ potentials support transitions beyond nearest neighbors,
see Fig. \ref{Fig2}c, thus the resulting population dynamics is rather complex.


\subsection{Cumulant Generating Function}
\label{B}

In molecular electronic applications we are prominently interested in the charge transport characteristics of the junction.
In order to ``count" charge transfer processes,   we define the so-called characteristic function \cite{BijayCGF,Esposito-QME}
\bea
\mathcal Z(\lambda_{e},\lambda_{p}) =
\langle e^{i\lambda_e \hat H_R + i\lambda_p \hat N_R}
e^{-i\lambda_e \hat H_R^H(t) - i\lambda_p \hat N_R^H(t)}\rangle,
\label{eq:Z1}
\eea
with $\lambda_e$ and $\lambda_p$ as counting fields for energy and particles, respectively,
transferred from the right terminal to the left one.
Operators here are written in the Heisenberg representation. The average is performed with respect to the total density matrix
(subsystem + baths) at
the initial time.   Equation (\ref{eq:Z1}) can be organized as
\bea
\mathcal Z(\lambda_e,\lambda_p)= {\rm Tr}_{S} \left[\rho_{\lambda_e,\lambda_p}^{S}(t)\right]
\label{eq:Z2}
\eea
with  the counting-fields dependent reduced density matrix 
\bea
\rho^{S}_{\lambda_e,\lambda_p}(t) \equiv
{\rm Tr}_{el,ph} \big[ \hat U_{-\lambda_e/2, -\lambda_p/2}(t)\, \rho_T(0)  \, \hat U^{\dagger}_{\lambda_e/2, \lambda_p/2}(t)\big].
\nonumber\\
\label{eq:vib}
\eea
The forward and backward evolution operators are {\it not} hermitian conjugates.
For example, the forward propagator is given by
\bea
&&\hat U_{-\lambda_e/2, -\lambda_p/2} (t)=
\nonumber\\
&& \exp\left[{-i \frac{\lambda_e}{2} \hat H_R - i \frac{\lambda_p}{2} \hat N_R}\right]\, {\hat U}(t) \,\exp\left[{i \frac{\lambda_e}{2} \hat H_R + i \frac{\lambda_p}{2} \hat N_R}\right]
\nonumber \\
&&\equiv \exp[-i \hat H_{-\lambda_e/2,-\lambda_p/2} (t)],
\eea
with the counting-field dependent total Hamiltonian, e.g.,
\bea
&&{\hat H}_{-\lambda_e/2,-\lambda_p/2} = \hat H_{S} +
\nonumber\\
&&  \hat H_{el} + \hat S \otimes \big[g \sum_{l,r} \gamma_{l}^{*} \gamma_r a_{l}^{\dagger} a_{r} e^{\frac{i}{2}
( \lambda_p +\epsilon_r \lambda_e)} + {\rm h.c.} \big] +
\nonumber\\
&&  \hat H_{ph} + \hat S\otimes \sum_k\nu_k\left(\hat b_k^{\dagger}+\hat b_k \right)
\label{eq:Hlam}
\eea
To evaluate the characteristic function we
therefore need to study the dynamics of the counting-field dependent reduced density matrix
with time evolution operators made of
the interaction Hamiltonian (\ref{eq:VelD})---now decorated with the counting-fields \cite{BijayPRB15} $\lambda=(\lambda_p,\lambda_e)$,
\bea
{\hat B}^{el}_{\mp \lambda/2}=g
\left[\gamma_{l}^* \gamma_r \hat a_{l}^{\dagger} \hat a_{r} e^{ \pm \frac{i}{2} (\lambda_p + \epsilon_r \lambda_e)} + h.c. \right].
\eea
We can now follow  standard weak-coupling projection operator methods,
work under the Markovian and secular approximations,
and receive an equation of motion for the counting-field dependent mode population \cite{BijayPRB15},
precisely analogous to Eq. (\ref{eq:pop}),
\bea
\dot{p}^{\lambda}_{n}(t)= -  p_n^{\lambda}(t) \sum_{m}  k_{n\to m}
+ \sum_{m}  k^{\lambda}_{m \to n}  {p}^{\lambda}_{m}(t).
\label{eq:poplambda}
\eea
%
The rate constants satisfy
\bea
k_{n\to m}^{\lambda}=\left[k_{n \to m}^{\lambda}\right]^{L \to R}
+ \left[k_{n \to m}^{\lambda}\right]^{R \to L} + k_{n\to m}^{ph},
\eea
recovering  Eq. (\ref{eq:ratessum}) when $\lambda=0$.
The counting-fields dependent terms are given by \cite{BijayPRB15,Bijay16}
\begin{widetext}
\bea
\big[ k^{\lambda}_{n \to m}\big]^{L \to R} &=& | S_{m,n}|^2\int_{-\infty}^{\infty}\frac{d\epsilon}{2\pi} f_L(\epsilon) (1-f_R(\epsilon+E_{nm})) J_L(\epsilon) J_R(\epsilon+E_{nm})
 e^{-i(\lambda_p + (\epsilon + E_{nm})\lambda_e)},
\nonumber\\
\big[k^{\lambda}_{n\to m}\big]^{R \to L} &=& |  S_{m,n}|^2
\int_{-\infty}^{\infty} \frac{d\epsilon}{2\pi} f_R(\epsilon) (1-f_L(\epsilon+E_{nm})) J_R(\epsilon) J_L(\epsilon+E_{nm})
 e^{i(\lambda_p+ \epsilon \lambda_e)}.
\label{eq:ratel2}
\eea
\end{widetext}
Obviously, the phonon bath-induced rates are intact in the present counting statistics calculation.
We can rationalize Eq. (\ref{eq:ratel2}) as follows:
According to our sign convention charge transferred is counted positive when flowing $R$ to $L$.
The rate $ [k^{\lambda}_{n \to m}\big]^{L \to R}$ stands for the process with a single electron crossing the junction {\it against}
this convention, adding an energy
in the amount of $\epsilon+E_{nm}$ to the $R$ bath.
The exponent, with charge and energy counting fields, therefore appears with a negative sign.
In contrast, the rate $ [k^{\lambda}_{n \to m}\big]^{R \to L}$ describes the transfer of an electron with energy
$\epsilon$ right-to-left, in line with our sign convention. The exponent then appears with a positive sign
decorating the counting fields.

It is convenient to organize Eq. (\ref{eq:poplambda}) as a matrix operation,
\bea
|\dot p^{\lambda}\rangle &=&{\cal L}^{\lambda} |p^{\lambda}\rangle
\nonumber\\
&=& \left({\cal L}^{\lambda}_{L \to R} + {\cal L}^{\lambda}_{R\to L} +   {\cal L}^{el}_{diag} +
 {\cal L}_{ph}\right) |p^{\lambda}\rangle.
\eea
${\cal L}^{el}_{diag}$ is a diagonal matrix with electronic bath relaxation rates,
independent of the counting field, see Eq. (\ref{eq:poplambda}).

Back to Eq. (\ref{eq:Z2}), the long-time (steady state) solution of Eq. (\ref{eq:poplambda}) hands over the
cumulant generating function (CGF),
\be
{\cal G}(\lambda) = \lim_{t \to \infty} \frac{1}{t} \ln {\cal Z} (\lambda)=
 \lim_{t \to \infty} \frac{1}{t} \ln \langle I|{  {p^{\lambda}}}(t)\rangle,
\label{eq:longCGF}
\ee
where $\langle I|=(1,1,1,\cdots)^T$ is the identity vector.
The CGF delivers the steady state charge and energy currents, as well as higher order cumulants,
by taking derivatives with respect to the counting fields [recall the definition Eq. (\ref{eq:Z1})].
%
%

\subsubsection{Charge current}
The charge current is derived from
\bea
\langle I_p \rangle &=& \frac{\partial {\cal G}(\lambda)}{\partial (i \lambda_p)}\Big|_{\lambda=0} =
\langle I |
\frac{\partial {\cal L}^{\lambda}}{\partial {(i\lambda_p)}}{\Big|}_{\lambda=0} p_{ss} \rangle
\label{eq:current-QME}
\eea
where $|p_{ss}\rangle=|p_0,p_1,p_2, \cdots \rangle$ is the column vector with the steady state populations,
obtained by solving  Eq. (\ref{eq:pop}),
$\dot p_n=0$, with the normalization condition $\sum_n p_n=1$.
We organize next working expressions for the charge current based on Eq. (\ref{eq:current-QME}).
First, one can immediately receive the intuitive construction
\bea
\langle I_p\rangle&=&
\sum_{m,n} p_n^{ss} \frac{\partial k_{n \to m}^{\lambda}}{\partial (i\lambda_p) } \Big|_{\lambda=0}
\nonumber\\
&=&
\sum_{m,n} p_n^{ss} \left( k_{n\to m}^{R\to L} - k_{n\to m}^{L \to R} \right),
\label{eq:curr1}
\eea
Another convenient form is based on the identification of the subsystem and bath correlation functions. In real time we define
\bea
C_S(\tau)&\equiv& \langle \hat S(0)\hat S(\tau) \rangle_{ss}
= \sum_n p_n^{ss} \langle n|  \hat S(0)\hat S(\tau)|n\rangle,
\eea
and
$C_{\nu,\bar \nu}(\tau) = \langle \hat B_{\nu}(0) \hat B_{\bar \nu}(\tau)\rangle$.
The frequency domain functions $C_S(\omega)$, $C_{\nu,\bar \nu}(\omega)$, are included in Appendix A.
We now organize the charge current as,
\bea
&&\langle I_p\rangle =\int_{-\infty}^{\infty}d\tau C_S(\tau) \left[ C_{RL}(\tau)-C_{LR}(\tau) \right]
\nonumber\\
&&= \frac{1}{2\pi}\int_{-\infty}^{\infty}d\omega\, C_S(-\omega) \left[ C_{RL}(\omega)-C_{LR}(\omega) \right].
\label{eq:curr2}
\eea
For more details, see Appendix A.

We emphasize that Eqs. (\ref{eq:current-QME}), (\ref{eq:curr1}), and (\ref{eq:curr2}) are equivalent:
One can compute the charge current directly from the Liouvillian, or by combining its matrix elements.
Alternatively, one can evaluate the correlation functions of the subsystem and the electronic baths in frequency domain,
in steady state,
and reach the charge current from their convolution.
Note that $p_{n}^{ss}$ depends on the coupling strength of the primary mode to both the electronic and phononic baths.

\subsubsection{Energy current}
The energy current is obtained from Eq. (\ref{eq:longCGF}) by taking the $\lambda_e$ derivative,
\bea
\langle I_e \rangle &=& \frac{\partial {\cal G}(\lambda)}{\partial (i \lambda_e)}\Big|_{\lambda=0}
= \langle I | \frac{\partial {\cal L}^{\lambda}}{\partial {(i\lambda_e)}}{\Big|}_{\lambda=0} p_{ss} \rangle.
\nonumber\\
&=&
\sum_{m,n} p_n^{ss} \frac{\partial k_{n \to m}^{\lambda}}{\partial (i\lambda_e) } \Big|_{\lambda=0},
\label{eq:Ecurrent-QME}
\eea
Defining the correlation functions $\dot C_{LR}(\tau) \equiv \langle \frac{d\hat B_{L}(0)}{dt} \hat B_{R}(\tau)\rangle$,
and $\dot C_{RL}(\tau) \equiv \langle \hat B_{R}(0)\frac{d\hat B_{L}(\tau)}{dt}\rangle$,
with $d\hat B_L/dt=i[\hat H_L,\hat B_L]$,
we can express the energy current as,
\bea
&&\langle I_e\rangle =\int_{-\infty}^{\infty}d\tau C_S(\tau) \left[ \dot C_{RL}(\tau)-\dot C_{LR}(\tau) \right]
\nonumber\\
&&= \frac{1}{2\pi}\int_{-\infty}^{\infty}d\omega \,\omega C_S(-\omega) \left[ C_{RL}(\omega)-C_{LR}(\omega) \right].
\label{eq:curre2}
\eea
For details, see Appendix A.
Eqs. (\ref{eq:curr2}) and (\ref{eq:curre2}) for the charge and energy current clearly portray the inelastic many-body nature of
transport processes in our model. Particles and energy transfer between the two metals proceed by the excitation/relaxation of the
subsystem oscillator.
These expressions also illustrate that our work only accounts for weak subsystem-bath coupling effects, as
multi-quanta effects are missing. As well, non-secular processes are non included.


\section{Simulations}
\label{result}

Considering the molecular junction setup of Fig. \ref{Fig1}, quantities of interest are the
current-voltage characteristics of the system (beyond linear response),
and its thermoelectric efficiency. 
In our simulations we assume metals with a constant density of states and a high energy cutoff.
For simplicity, we consider a symmetric setup with
$\Gamma=\Gamma_L=\Gamma_R$ and $\epsilon_0=\epsilon_{d,a}$.
Recall that the functions $J_{\nu}(\epsilon)$
describe the density of states in the $\nu$ compartment---after absorbing
the molecular electronic levels into the metal leads (e.g., the donor state into the $L$ metal).
%

\begin{figure*} [t]
\hspace{4mm}
\includegraphics[width=15cm]{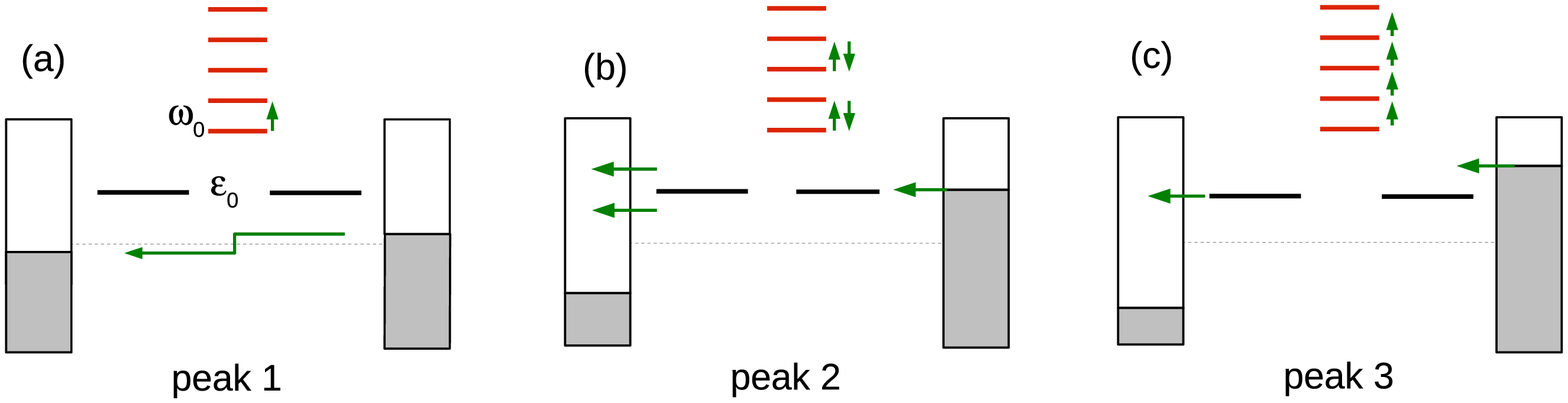}
\vspace{-70mm}
\caption{
Harmonic mode junction.
Illustrations of principal processes that contribute to the three peaks in the differential conductance
of Fig. \ref{IVHO}.
(a) Low-bias excitations satisfying $\Delta \mu=\omega_0$ are responsible for peak 1.
(b) A resonant condition is met once $\Delta\mu=2\epsilon_0$, leading to peak 2.
(c)  When $\Delta\mu=2(\epsilon_0 +\omega_0)$, a vigorous mode-heating mechanism generates peak 3.
The dashed line marks the equilibrium Fermi energy. Horizontal green arrows represent incoming and outgoing
electrons of different energies. Vertical arrows exemplify corresponding vibrational relaxation and excitation processes.}
\label{mechHO}
\end{figure*}

\begin{figure*}
\hspace{-4mm}
\includegraphics[width=9cm]{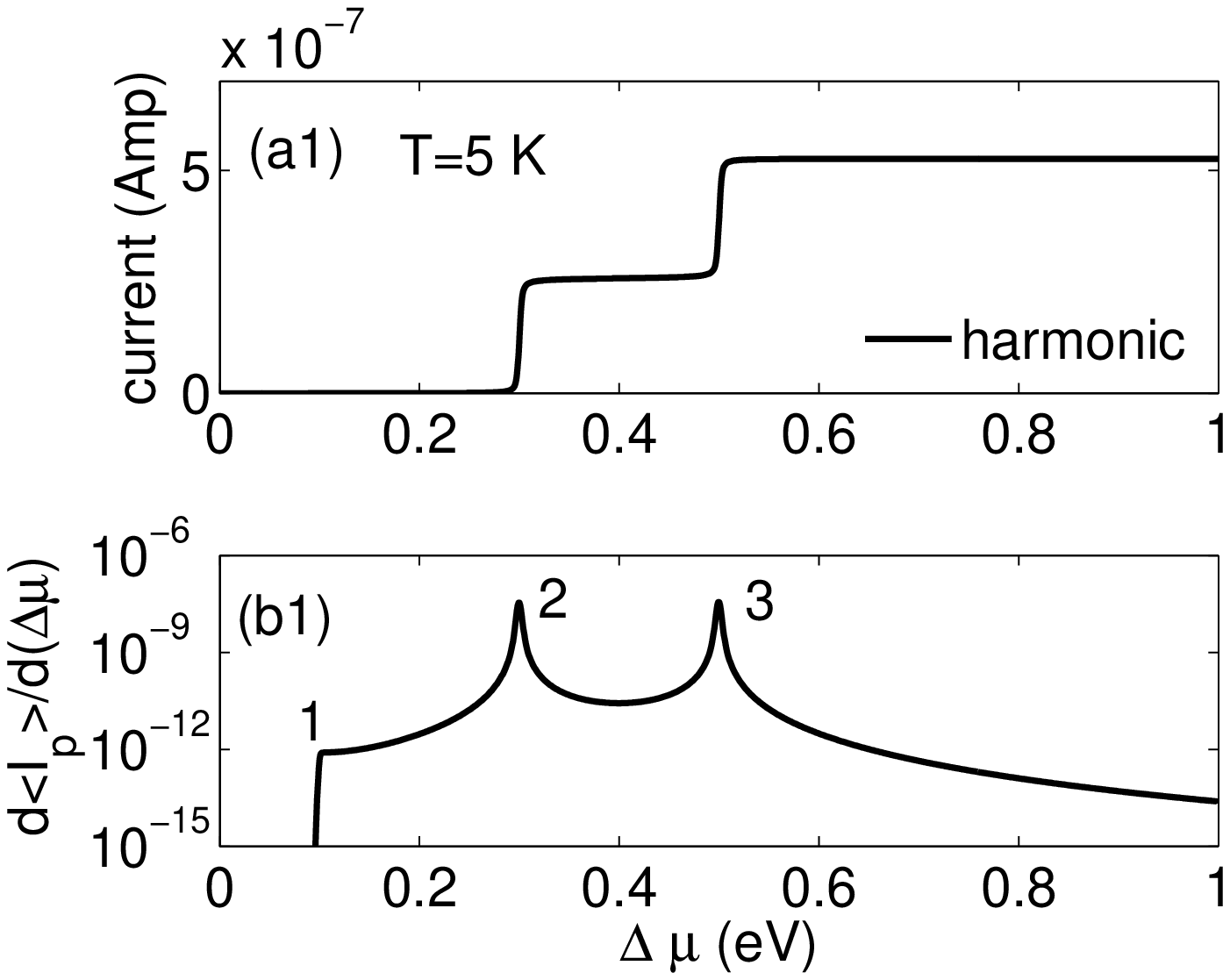}
\includegraphics[width=9cm]{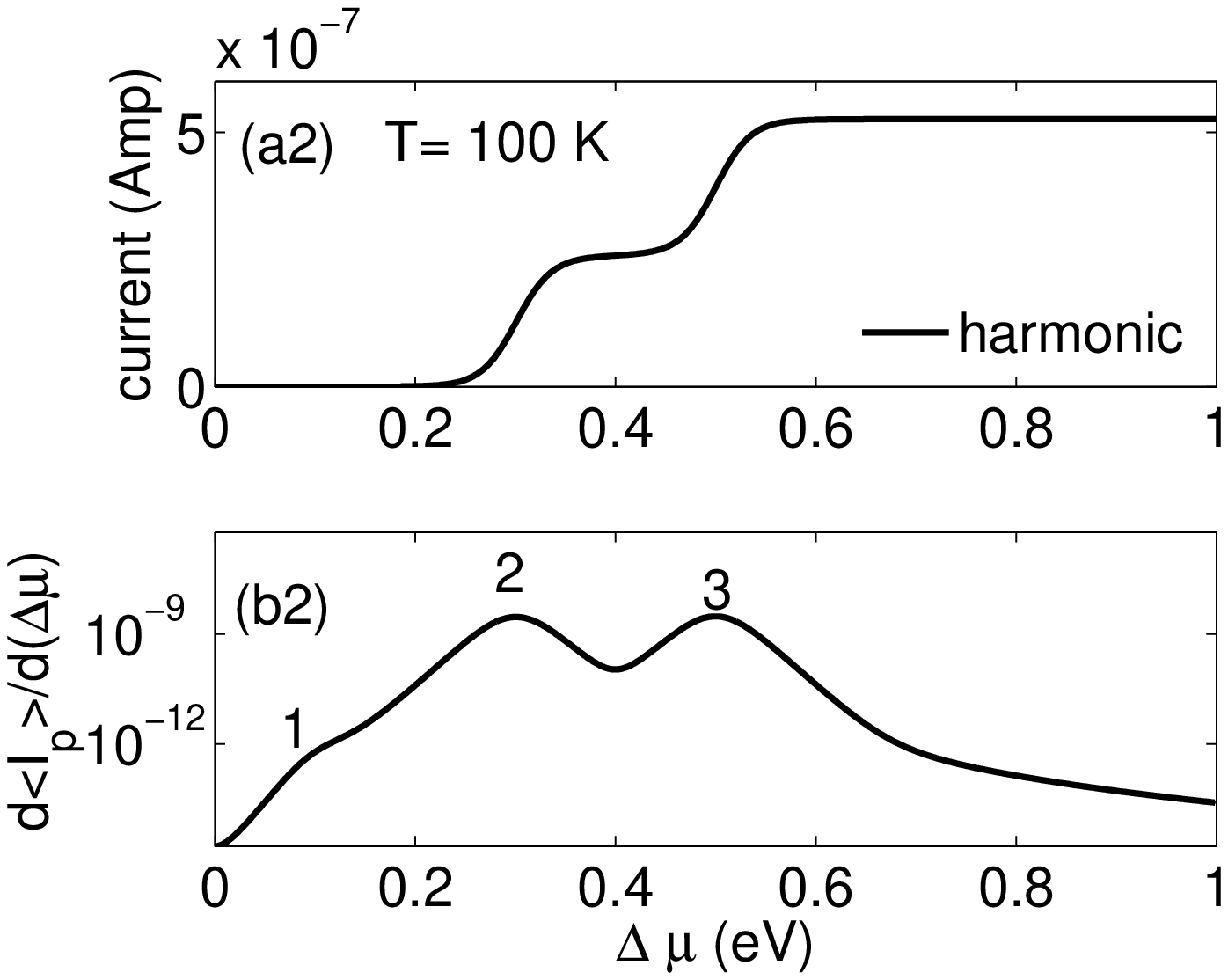}
\caption{
Harmonic mode junction.
(a) Current-voltage characteristics ($\langle I_p \rangle$ as a function of $\Delta \mu$)
and (b) differential conductance at
$T=5$ K (left)  and at a higher temperature, $T=100$ K (right).
The differential conductance exposes three peaks--- the corresponding inelastic processes are
illustrated in Fig. \ref{mechHO}.
Parameters are $\epsilon_0=0.15$, $\omega_0=0.1$, $g=0.1$, $\Gamma=0.001$, $\Gamma_{ph}=0.05$ in eV.
}
\label{IVHO}
\end{figure*}

\begin{figure*} [pt]
\hspace{4mm}
\includegraphics[width=15cm]{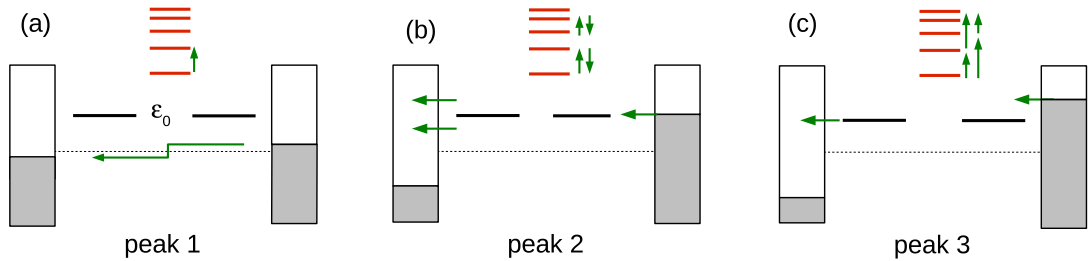}
\vspace{-70mm}
\caption{
Anharmonic-potential junction.
Illustrations of principal processes contributing to the differential conductance in e.g., the Morse potential junction, examined in Fig. \ref{IVAH1}.
(a) Low-bias excitation, satisfying $\Delta \mu =E_1-E_0$, responsible for peak 1 in Fig. \ref{IVAH1}.
(b) Resonant conduction $\Delta\mu=2\epsilon_0$, leading to peak 2, and
(c) mode-heating regime with $\Delta\mu=2(\epsilon_0 +E_{n}-E_m)$, $n>m$, generating peak 3.
The dashed line marks the equilibrium Fermi energy. Horizontal green arrows represent incoming and outgoing electrons of different
energies, vertical arrows exemplify vibrational excitations.
}
\label{mechAH}
\end{figure*}

\begin{figure*}
\hspace{-8mm}
\includegraphics[width=18cm]{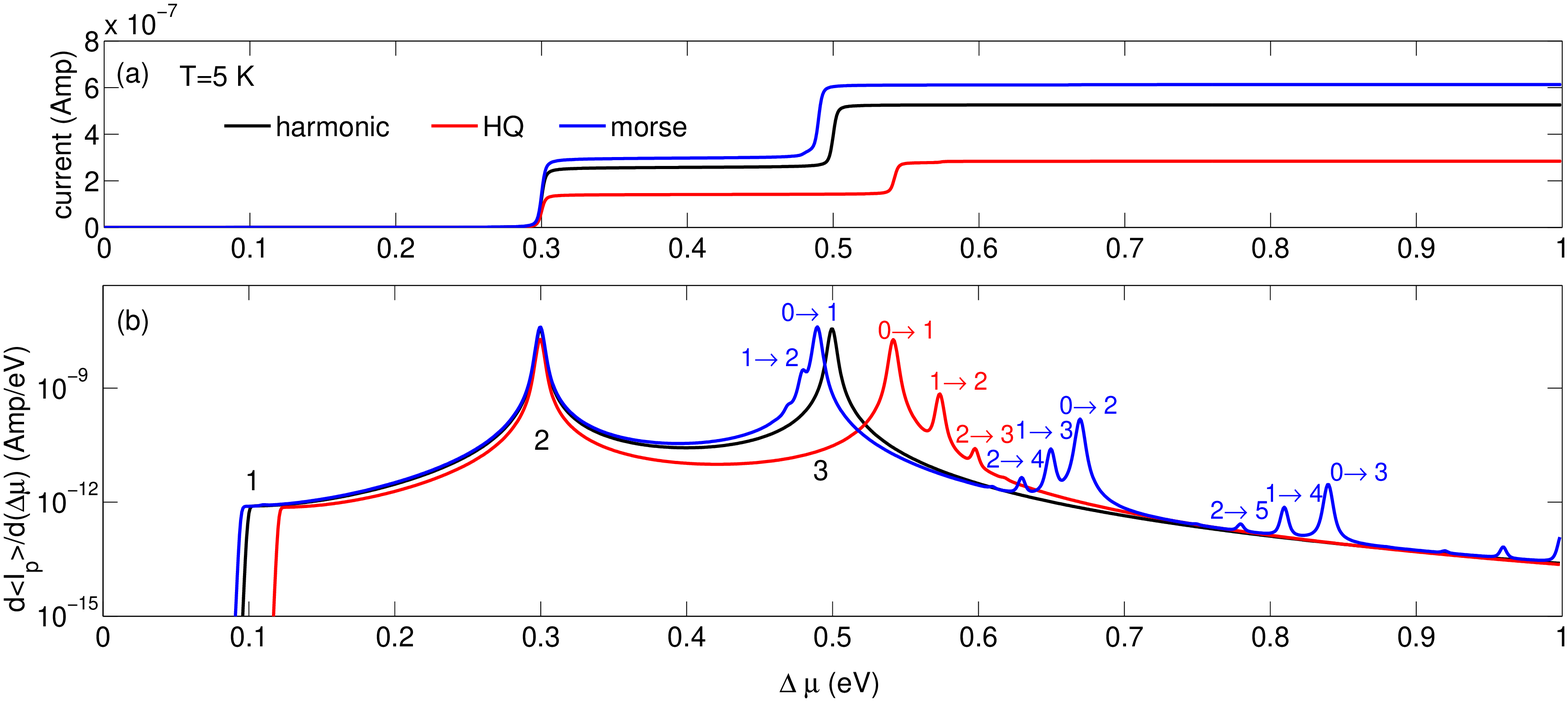}
\caption{(a) Current-voltage characteristics ($\langle I_p\rangle$ as a function of $\Delta\mu$)
 and (b) differential conductance of the DA junction
with harmonic (black) Morse (blue) and the HQ (red) oscillators at low temperature, $T=5$ K.
Parameters are 
$\epsilon_0=0.15$, $\omega_0=0.05$, $g=0.1$, $\Gamma=0.001$, $\Gamma_{ph}=0.05$ in eV, $T=5$ K.
Anharmonicity parameters for the HQ and the Morse potentials are $a_4=1$ 1/eV and $D=1$ eV, respectively.
}
\label{IVAH1}
\end{figure*}

\begin{figure}
\hspace{-8mm}
\includegraphics[width=8cm]{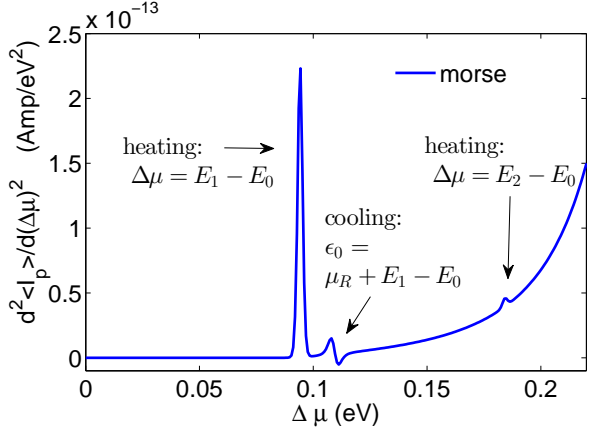}
\caption{Second derivative of the current---presented in Fig \ref{IVAH1}(a)---with respect to voltage,
considering a junction with a Morse oscillator.
Parameters are the same as in Fig. \ref{IVAH1}.
}
\label{Diff2}
\end{figure}

\begin{figure*}
\hspace{-8mm}
\includegraphics[width=7cm]{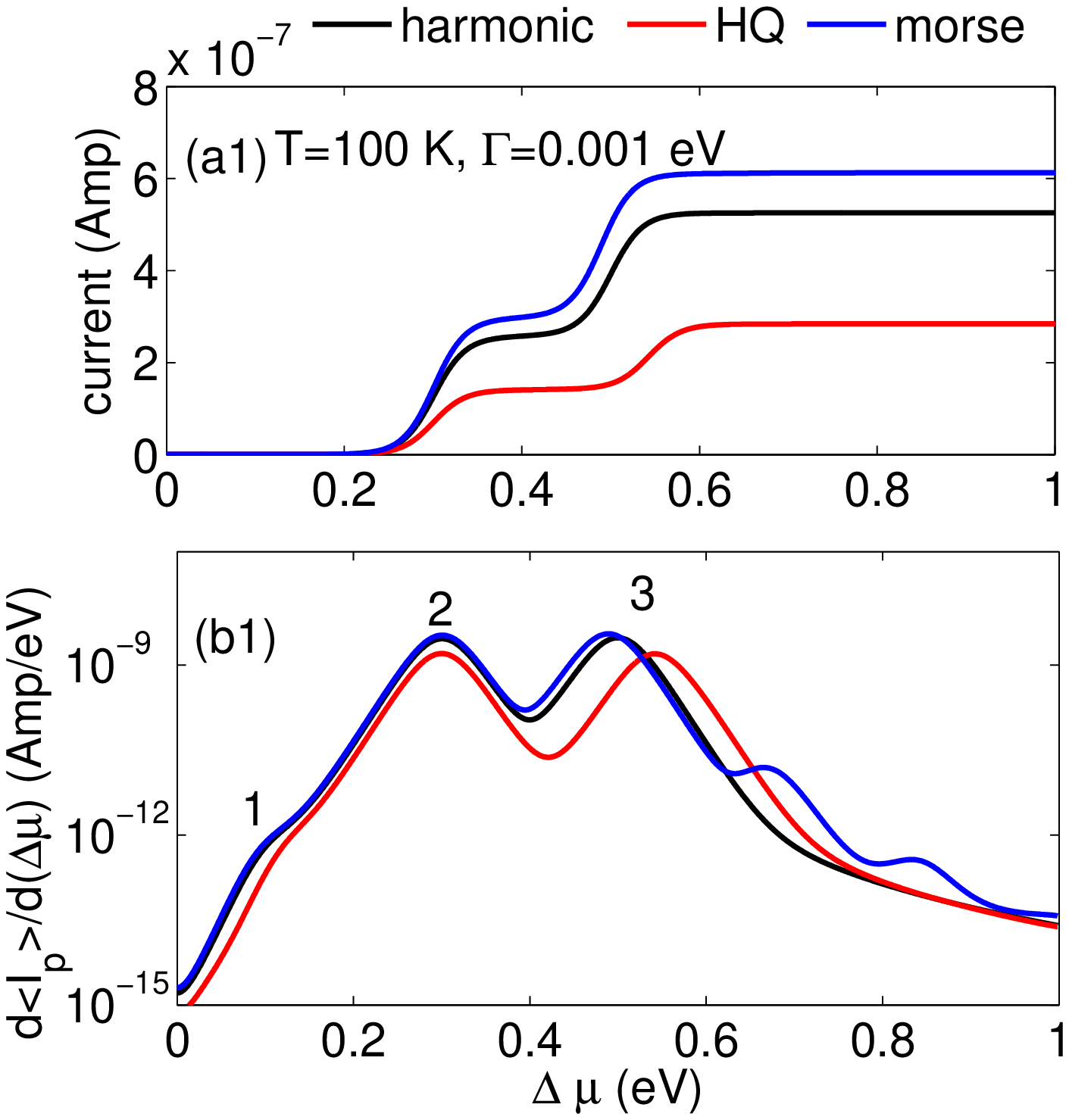}
\hspace{5mm}
\includegraphics[width=7cm]{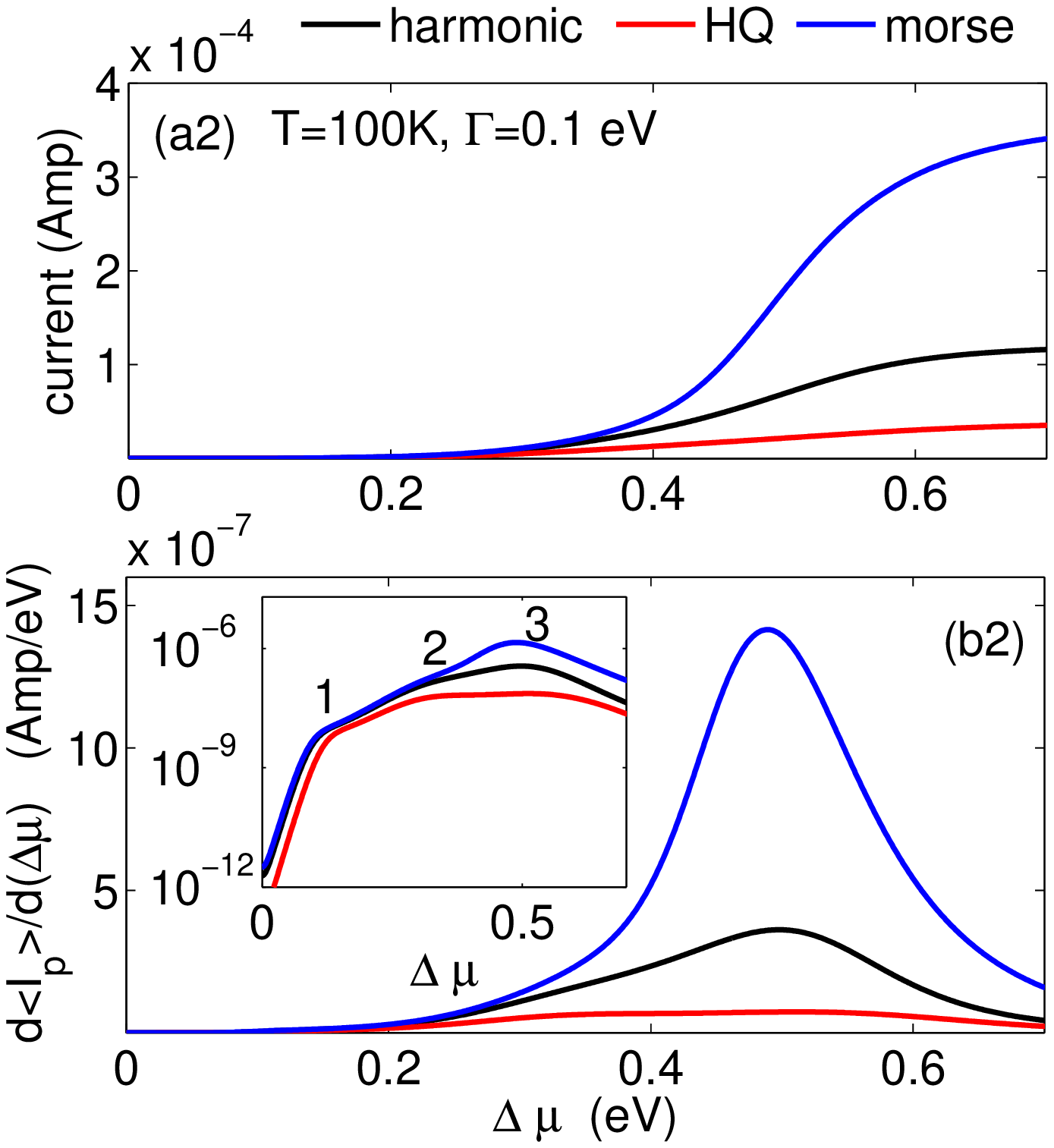}
\caption{(a) Current-voltage characteristics and (b) differential conductance of the DA junction
with harmonic (black) Morse (blue) and an HQ (red) oscillators at $T=100$ K with weak (left) and moderate
(right) metal-molecule hybridization $\Gamma$.
Other parameters are the same as in Fig. \ref{IVAH1}.
}
\label{IVAH2}
\end{figure*}


\subsection{Charge current-voltage characteristics}

We set the equilibrium Fermi energy at zero and apply the voltage bias in a symmetric manner, $\mu_R=-\mu_L>0$.
According to our sign convention, the charge current is positive when flowing right to left.
We assume that the molecular orbitals do not shift with bias.
This effect could be implemented easily to materialize a strong diode behavior
\cite{SiminePCCP,SimineINFPI}.

The main question that we address next
concerns signatures of the anharmonic molecular oscillator on
the charge current. Recall that our model only supports inelastic (vibrationally-assisted)
electron current.
In Appendix B  we further add a direct tunneling term between the two sites to the Hamiltonian,
 $t\hat c_d^{\dagger}\hat c_a$ +h.c. This elastic contribution to the current is included  (as an approximation)
by the coherent Landauer formula on top of the inelastic contribution.

\subsubsection{Harmonic molecular oscillator}
We begin by studying transport behavior in the harmonic-mode junction.
Inelastic scattering mechanisms are illustrated in Fig. \ref{mechHO}; the current-voltage characteristics and the differential conductance,
at two different temperatures, are depicted in Fig. \ref{IVHO}.
We use $\epsilon_0=0.15$, $\omega_0=0.1$, $\Gamma$=0.001, $\Gamma_{ph}=0.05$, all in eV, and $T$=5 K and $T=100$ K.
This choice of parameters allows us to resolve three peaks in the differential conductance, and we now explain these features.

The lowest peak (1) in Fig. \ref{IVHO}(b) appears around $\Delta \mu=0.1$ eV, once electrons acquire sufficient energy
to be exchanged with the vibrational mode---of frequency $\omega_0=0.1$ eV.
Nevertheless, the current is very small at this region since the molecular
electronic levels are positioned outside the bias window, $\epsilon_0>\Delta\mu/2$.

The second (2) peak in the differential conductance arises around $\Delta \mu\sim 2\epsilon_0$,
once a resonant condition is met,
with the chemical potential at the right lead reaching the energy of the (degenerate) molecular orbitals.
Outgoing electrons at the left lead emerge from the junction with energies around $\epsilon_0\pm \omega_0$,
with the plus (minus) sign corresponding to relaxation (excitation) processes of the vibrational mode.

The upper peak (3) in the differential conductance develops around $\mu_R\sim (\epsilon_0+\omega_0)$.
At this bias, incoming electrons---of energies $\epsilon_0+\omega_0$---
excite the vibrational mode,  giving away $\omega_0$
and leaving the junction with energy $\epsilon_0$, in a region of high density of states;
recall that $J(\epsilon)$ shows a maximum at $\epsilon_0$.
This peak in the differential conductance thus principally corresponds to heating effects of the vibrational mode,
processes that can be contained by allowing energy dissipation from the primary mode to a secondary phonon bath,
using $\Gamma_{ph}\neq0$.

\subsubsection{Anharmonic molecular oscillators}
We proceed and examine the role of potential anharmonicity on the current and the differential conductance.
Fig. \ref{mechAH} depicts relevant inelastic mechanisms. Figs. \ref{IVAH1}-\ref{IVAH2} display
the current-voltage characteristics and the differential conductance
at different temperatures and metal-molecule hybridization.

The low temperature weak-hybridization behavior of an anharmonic-mode junction is displayed in Fig. \ref{IVAH1}.
We can readily identify the first peak (1) in Fig. \ref{IVAH1} (compare to Fig. \ref{IVHO}) by the sharp vertical jump in the differential conductance
around $\Delta\mu=0.1$ eV.
The precise position of the peak depends on the nature of the potential.
In contrast, the position of the second peak (2) in the differential conductance is not affected by the nature of the vibrational
potential---it is determined by a resonant condition for the electronic system, $\mu_R=\epsilon_0$.
The third peak (3) is largely influenced by the potential anharmonicity. Particularly for
the Morse potential, the peak is split and replicated at high voltage as we explain next.

We identify three central effects of anharmonicity on conductance:
(i) Magnitude of current. The Morse (HQ) potential supports the highest (lowest) currents.
(ii) Shift of peaks.
The first and third peaks are red (blue)  shifted for the Morse (HQ) model relative to the harmonic oscillator case.
(iii) Splitting of the third peak and appearance of new peaks at high bias.
In the examined range of bias, the Morse potential supports succession of peaks at high bias.
These peaks are missing altogether in the HO model.

Observations (i)-(ii) can be reasoned by recalling the role of anharmonicity on level spacing:
Energy levels in the HQ potential become further apart as we go higher in energy, with spacings exceeding the harmonic value $\omega_0$.
In contrast, in the Morse potential levels are pushed together, see Fig. \ref{Fig2}.
These adjustments to level spacings shift the location of the first and third peaks.
More significantly, when energy levels cluster, heating processes become more feasible, enhancing the current
at high bias.

We now explain observation (iii).
The third peak in the differential conductance emerges due to heating effects of the molecular vibration.
In harmonic modes only transitions between neighboring levels are allowed and
gaps between levels are fixed. This translates to a {\it single} peak at $\Delta \mu = 2(\epsilon_0+\omega_0)$.
The HQ and the Morse potentials, in contrast, support energy spectrum with varying energy spacings---leading to the
splitting of the third peak. This splitting is particularly significant for the HQ model;
the transitions $|n\rangle\to |n+1\rangle$
can be readily resolved at $\Delta\mu=2(\epsilon_0+E_{n+1}-E_n)$, see e.g. the peaks at 0.54,0.57, 0.597 eV.
Anharmonic potentials further
relax the strict harmonic ``selection rule", allowing transitions beyond nearest-neighboring states.
Specifically, the excitations $|n\rangle\to |n+2 \rangle$ are allowed for the Morse potential,
showing up as a succession of three peaks
for $|2\rangle\to |4\rangle$, $|1 \rangle\to |3\rangle$ and $|0\rangle\to |2\rangle$, from low to high frequencies.
These transitions are strictly forbidden for the HO and the HQ potentials given the even symmetry of the potential.

We now more carefully analyze the low-bias regime where peak (1) shows up,
by studying the second derivative of the current with respect to bias,
see Fig. \ref{Diff2}.
This type of analysis, inelastic electron tunneling spectroscopy (IETS) \cite{IETS1,IETS2,IETSrev}, has been demonstrated to
provide fundamental microscopic information on electron-vibration coupling in transport experiments, see e.g. Refs. \cite{GalpIETS04,GalpIETS04N,Zant15}.
For simplicity, we only analyze here the Morse potential.
We resolve three peaks in the second derivative, corresponding to different low-bias resonance situations.
The dominant low-bias effect is a heating process of the vibration, taking place at $\Delta \mu=E_1-E_0\sim 0.095$ eV.
Less likely yet visible are heating effects due to direct transitions
from the ground state to the {\it second} excited state satisfying  $\Delta \mu=E_{2}-E_0\sim 0.185$ eV.
In between, when the condition $\mu_R=\epsilon_0-E_{1}-E_{0}$  is reached,
the vibration is cooled down, and electrons gain sufficient energy so as to satisfy an electronic resonance condition and effectively cross the junction.
Within the present parameters for the Morse potential, this cooling situation is fulfilled at $\mu_R\sim 0.055$ eV, or $\Delta \mu=0.11$ eV.
Since temperature is rather low, this cooling process is quite limited compared to heating effects.
Note as well that within our choice of parameters, in the case of a harmonic oscillator, the heating $\Delta \mu=\omega_0$ and cooling
$\mu_R=\epsilon_0-\omega_0$ conditions (accidentally) coincide at $\Delta\mu=0.1$ eV.
However, since heating effects greatly dominate over cooling processes at low bias and low temperatures,
we had attributed above (Figs. \ref{mechHO} and \ref{mechAH}) the first peak to mode-heating effects.


Finally, we comment that the trends observed in Figs. \ref{IVAH1}-\ref{Diff2} are maintained at room temperature or at higher hybridization.
However, the separation between the different peaks becomes rather poor then, see Fig. \ref{IVAH2}.

\subsubsection{Asymptotic high-bias results}

An immediate observation from Figs. \ref{IVAH1} and \ref{IVAH2} is that the inelastic current is the highest for a junction with
a Morse mode, and the lowest for the HQ case.
We justify this observation by studying the behavior of the current in the high bias regime,
when the current is approximately uni-directional with electrons flowing right-to-left. 
Our starting point is equation (\ref{eq:curr2}) for the charge current,
\bea
\langle I_p\rangle =\frac{1}{2\pi}\int_{-\infty}^{\infty} d\omega\, C_{S}(\omega)\left[ C_{RL}(-\omega)-C_{LR}(-\omega)\right].
\eea
At low temperatures and in the high bias limit $\Delta \mu > T,\omega_s$,
with $\omega_s$ a characteristic frequency of the oscillator, the electronic correlation functions reduce to
\bea
C_{RL}(\omega)&=&
\frac{1}{2\pi} \int_{-\infty}^{\infty} d\epsilon f_R(\epsilon)\left[1-f_L(\epsilon-\omega) \right]J_R(\epsilon)J_L(\epsilon-\omega)
\nonumber\\
&\rightarrow&
\frac{1}{2\pi} \int_{\mu_L+\omega}^{\mu_R} d\epsilon  J_R(\epsilon)J_L(\epsilon-\omega)
\nonumber\\
C_{LR}(\omega)&=&0.
\eea
We assume that the hybridization is large, $\Gamma_{L,R}>\epsilon_{d,a}$ and receive from Eq. (\ref{eq:spec})
$J_{L,R}(\epsilon)=\frac{4g}{\Gamma_{L,R}}$. The charge current now simplifies to,
\bea
\langle I_p \rangle \approx \frac{1}{(2\pi)^2}  \frac{16g^2}{\Gamma_L\Gamma_R} \Delta \mu
\int_{-\infty}^{\infty} d\omega C_S(\omega).
\eea
We identify the system correlation function, evaluated as an expectation value over the steady-state solution, by
\bea
&&\frac{1}{2\pi}\int_{-\infty}^{\infty} d\omega C_S(\omega)= \sum_{m,n} p_n^{ss} |\langle n|\hat S|m\rangle|^2
\nonumber\\
&&=\langle \hat S^2(0)\rangle_{ss}.
\eea
This function depends on the voltage bias since the steady state populations of the oscillator
are obviously influenced by the electronic bath.
It describes the mean-square displacement of the oscillator, in steady state.
We can now organize a rather compelling expression for the  inelastic current,
\bea
\langle I_p \rangle
\approx \frac{1}{2\pi}\frac{16g^2}{\Gamma_L\Gamma_R} \Delta \mu \langle \hat S^2(0) \rangle_{ss}.
\eea
It grows with the electron-oscillator coupling strength as  $g^2$,
and it depends on the electronic hybridization as  $(\Gamma_L\Gamma_R)^{-1}$,
with $\Gamma_{L,R}^{-1}$ as the lifetime of electrons in the donor/acceptor states.
Furthermore, the scaling with the
 mean-square displacement demonstrates that oscillators with  a highly confined motion (e.g., the HQ potential),
support low currents relative to softer oscillators (e.g., the Morse potential).

How does the charge current scale with $\Delta\mu$?
For the HO case we readily calculate the mean square displacement at an arbitrary voltage.
Following Ref. \cite{Bijay16} we obtain
\bea
\langle \hat S^2(0)\rangle_{ss} = \frac{k_d+k_u}{k_d-k_u},
\eea
with $k_d$ and $k_u$ as the relaxation and excitation rate constants.
Neglecting phonon relaxation rates (assuming an isolated primary mode),
it can be shown that
$k_d+k_u=\frac{16g^2}{\pi\Gamma_L\Gamma_R}\Delta\mu$,
$k_d-k_u=\frac{16g^2}{\pi\Gamma_L\Gamma_R}\omega_0$.
thus, $\langle \hat S^2(0)\rangle_{ss}=\Delta\mu/\omega_0$, and
the charge current  obeys a quadratic relation--at high bias,
\bea
\langle I_p \rangle \approx  \frac{8g^2}{\pi\Gamma_L\Gamma_R} \frac{\Delta\mu^2}{\omega_0}.
\eea
We emphasize that this scaling was derived for a molecular junction with harmonic nuclear motion.
It describes the current-voltage characteristics at
high bias, $\Delta\mu>\omega_0,T$, strong hybridization $\Gamma_{L,R}>\epsilon_{d,a}$, and for an isolated mode, $\Gamma_{ph}=0$.
Fig. \ref{IVAH2}(a2) was generated with parameters outside this restrictive region, yet
we observe that the three cases, HO, HQ and Morse, display a quadratic scaling $\langle I_p\rangle\propto \Delta\mu^2$
at intermediate biases; at very high bias $\Gamma_{ph}$ is responsible for the saturation behavior.

\begin{figure}
\hspace{-4mm}
\includegraphics[width=15cm]{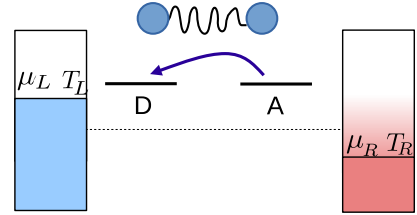}
\vspace{-80mm}
\caption{DA junction under voltage bias and temperature difference. In our calculations
the right (left) terminal is made hot (cold). The
thermoelectric energy conversion efficiency is defined in Eq. (\ref{eq:eff}).
}
\label{Seff}
\end{figure}

\begin{figure*}
\hspace{-4mm}
\includegraphics[width=15cm]{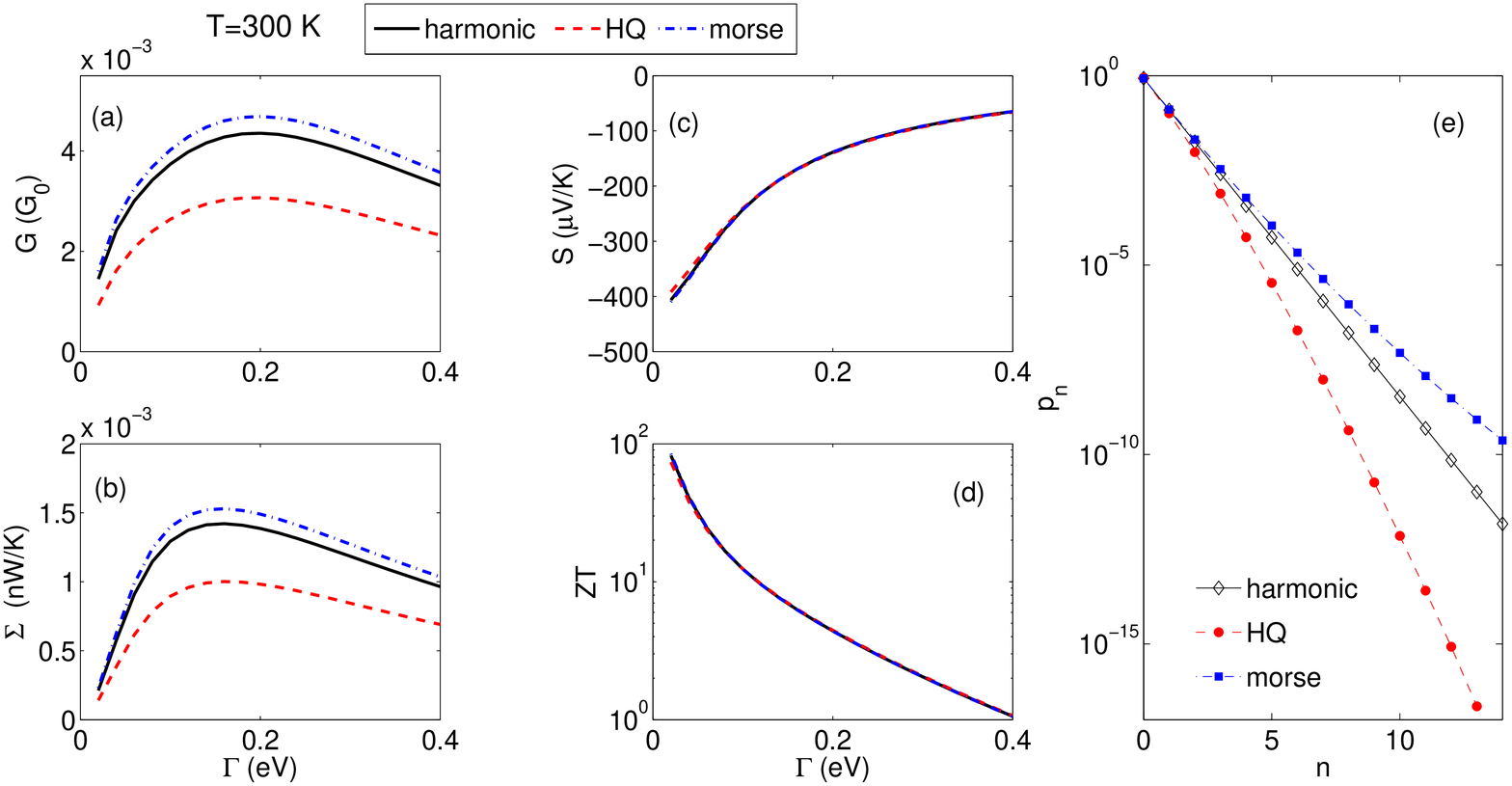}
\vspace{-4mm}
\caption{
Linear response behavior of the donor-acceptor junction as a function of the molecule-metal hybridization energy
at room temperature $T=$300 K 
with a harmonic mode (full), HQ (dashed), and the Morse  mode (dashed-dotted).
(a) Electrical conductance $G$ in units $G_0=e^2/h$ the quantum of conductance per channel per spin.
(b) Electronic thermal conductance $\Sigma$,
(c) Seebeck efficiency $S$,
and (d) figure of merit $ZT$.
(e) Population of vibrational states in the three models (independent of $\Gamma$).
Parameters are $\epsilon_0=0.15$, $\omega_0=0.05$, $g=0.01$ in eV, and temperature $T=300$ K.
Anharmonicity parameters for the HQ and the Morse potentials are $a_4=1$ 1/eV and $D=1$ eV, respectively.
}
\label{LR1}
\end{figure*}

\begin{figure*}
\hspace{2mm}
\includegraphics[width=17cm]{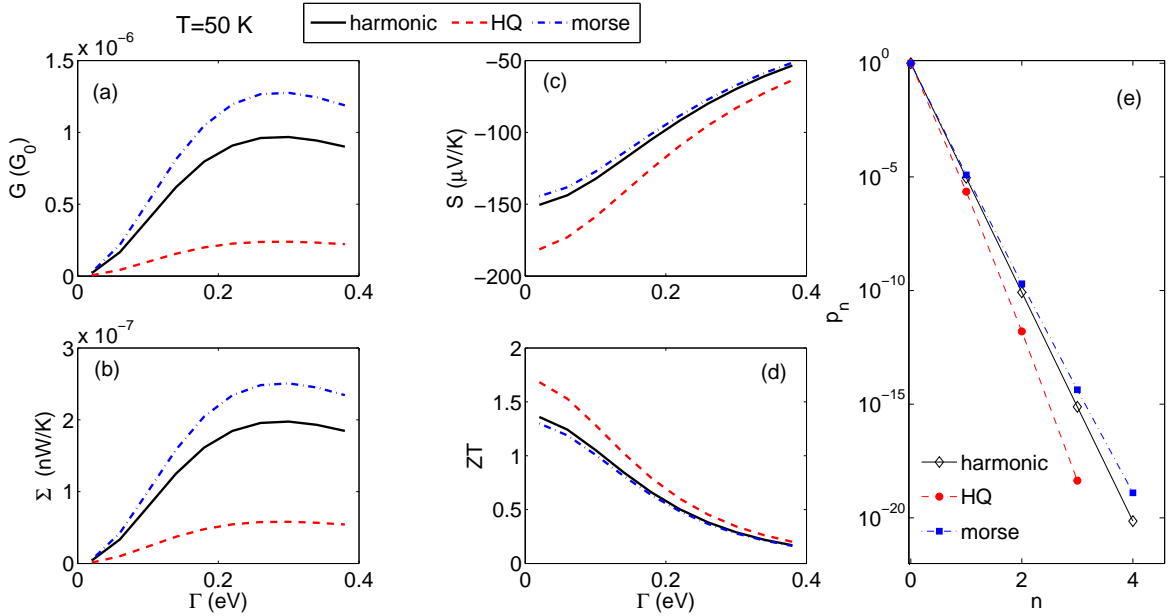}  
\caption{Linear response characteristics of the junction at low temperatures, $T=50$ K.
(a) Electrical conductance,
(b) Electronic thermal conductance,
(c) Seebeck coefficient,
and (d) figure of merit $ZT$.
Parameters are the same as Fig. \ref{LR1}.
}
\label{LR2}
\end{figure*}

\subsection{Thermopower and energy conversion efficiency}

In this Section we aim in identifying signatures of molecular anharmonicity in the thermopower and the
energy conversion efficiency.
To operate the device as a thermoelectric engine, we set $T_L<T_R$ and $\mu_L>\mu_R$.
We also isolate the oscillator from the secondary phonon bath so as heat dissipation is only permitted in the metals,
for a schematic representation, see Fig. \ref{Seff}.
Three-terminal engines were studied e.g. in Refs. \cite{Jiang1,JiangRev}.

The thermoelectric efficiency is
defined as the ratio between the averaged power generated by the engine and the heat absorbed from the hot (right) reservoir,
\bea
\eta= \frac{(\mu_L-\mu_R)\langle I_p\rangle}{\langle I_q\rangle},
\label{eq:eff}
\eea
with
$\langle I_q\rangle = \langle I_e\rangle - \mu_R\langle I_p \rangle$ as the heat current.
The linear-response and the nonlinear performance of the DA molecular junction were recently examined
in Ref. \cite{BijayDABeil}---considering either a harmonic mode, or a two-state system serving as
an anharmonic impurity.
We found there that the electrical and thermal conductances were sensitive to whether the mode was harmonic/two-state system.
However, we proved, based on the analytical form of the CGF, that
the Seebeck coefficient, the thermoelectric figure-of-merit, and the thermoelectric efficiency beyond
linear response, concealed this information. 
We now examine whether this insensitivity of the thermoelectric figure-of-merit to mode properties (harmonicity/anharmonicity)
is a general feature valid beyond the particular (and somewhat unique) two-state impurity case.

We begin our analysis with linear response coefficients, expanding the charge and heat current
around thermal equilibrium, with $\Delta V$ and $\Delta T$ as the voltage and temperature differences, respectively,
\bea
\langle I_p\rangle &=&G \Delta V + GS \Delta T
\nonumber\\
\langle I_q\rangle &=&G \Pi \Delta V + (\Sigma S \Pi)\Delta  T.
\eea
Here, $G$ is the electronic conductance,  
$S$ the thermopower (not to be confused with the subsystem operator $\hat S$),
$\Pi$ the Peltier coefficient, and $\Sigma$ the electric thermal conductance.
The (dimensionless) figure of merit  $ZT=\frac{GS^2}{\Sigma} T$ determines the
(linear response) thermoelectric energy conversion efficiency.

Representative results are displayed in Figs. \ref{LR1}-\ref{LR2}, where we
study the behavior of linear response coefficients as
a function of the metal-molecule hybridization at two different temperatures.
In agreement with Figs. \ref{IVAH1}-\ref{IVAH2},
we find  that the three models, harmonic, Morse, and HQ,
support distinct (electrical, thermal) conductances,
with the Morse potential junction showing the highest current and the HQ model demonstrating current suppression.
In contrast, the Seebeck coefficient and the figure of merit
in panels (c) and (d) display little sensitivity to mode anharmonicity:
At high temperatures  [quantified below Eq. (\ref{eq:curr2ee})] $S$ and $ZT$ are almost identical in the different models, with
about $5\%$ deviations. At low temperatures
and weak hybridization more substantial deviations show up,
with the HQ model allowing 20$\%$ higher thermoelectric efficiency than the Morse oscillator.

We further present in panel (e) of Figs. \ref{LR1}-\ref{LR2}
the long-time population of the vibrational state as a function of the level index $n$.
Note that the steady state population does not depend on
the coupling $\Gamma$ close-to-equilibrium.
We find that at the considered temperatures, $T=50-300$ K,
level occupation quickly drops with $n$, thus
charge transfer dynamics is essentially determined by transitions between the first two states.
We recall from previous work \cite{BijayDABeil} that in our junction---when assuming a two-state impurity mode---
the following trends are observed:
with increasing frequency $\omega_0$,
the electric and thermal conductances drop, the magnitude of $S$ grows, and $ZT$ increases.
This behavior precisely matches the enhancement of $ZT$ in the HQ model relative to the Morse case.

We now explain the high-temperature and large-$\Gamma$ insensitivity of $S$ and $ZT$,  quantities
which depend on ratio of currents, to the nature of the oscillator.
We begin  with Eq. (\ref{eq:curre2}) for the energy current, included here again for convenience,
\bea
\langle I_e\rangle
= \frac{1}{2\pi}\int_{-\infty}^{\infty}d\omega \omega C_S(-\omega) \left[ C_{RL}(\omega)-C_{LR}(\omega) \right].
\label{eq:curr2ee}
\eea
The function $C_S(\omega)=2\pi \sum_{n,m} p_n^{ss} |\langle m| \hat S|n\rangle|^2 \delta(\omega+E_{mn})$
depends on the nature of the oscillator. The electronic bath correlation functions,
e.g. $C_{RL}(\omega)= \frac{1}{2\pi} \int d\epsilon f_R(\epsilon)\left[1-f_L(\epsilon-\omega) \right]J_R(\epsilon)J_L(\epsilon-\omega)$,
are calculated at the oscillator transition frequencies $E_{mn}$.
Now imagine that $\Gamma$ is very small, to be quantified next.
The electronic spectral density functions $J_{L,R}(\epsilon)$ become then very narrow.
As a result, the convolution in Eq. (\ref{eq:curr2ee}) delicately depends on the level spacing supported by the oscillator.
In contrast, at relatively large $\Gamma$ and $T$,
$C_{\nu,\bar\nu}(\omega)$ maintains comparable values for a range of frequencies
$\omega_0-\delta\omega_0<\omega<\omega_0+\delta \omega_0$
with $\delta \omega_0 <\Gamma, T$.
Here, $\delta\omega_0$ is a measure for deviations from the harmonic  energy spacing $\omega_0$.
Within our parameters,
 $\delta \omega_0\sim \omega_0^2/D\sim 2$ meV
for the Morse potential while for the HQ oscillator, $\delta\omega_0\sim 20$ meV.
The insensitivity of $C_{\nu,\bar \nu}(\omega)$ to the precise value of the energy level spacings $E_{mn}$
allows us to approximate $\omega\rightarrow \omega_0$ in the integrand of Eq. (\ref{eq:curr2ee}),
making the energy current proportional to $\langle I_p\rangle$.
We conclude that as long as $T,\Gamma>\delta \omega_0$,
ratio of currents turn out independent of $C_{S}(\omega)$--- thus $S$ and $ZT$ become identical in harmonic and anharmonic junctions.
This statement is valid assuming that
currents are determined by the population of the lowest few states of the oscillator.

In agreement with this argument, Figures \ref{LR1}-\ref{LR2} demonstrate that in the harmonic and Morse potentials,
$S$ and $ZT$ are almost indistinguishable. In contrast, the HQ model deviates from the harmonic limit for these quantities
at $T=50$ K, translating to $k_BT=4$ meV, which is below  $\delta \omega_0= 20$ meV. 

\begin{figure*}[htbp]
\hspace{-4mm}
\includegraphics[width=15cm]{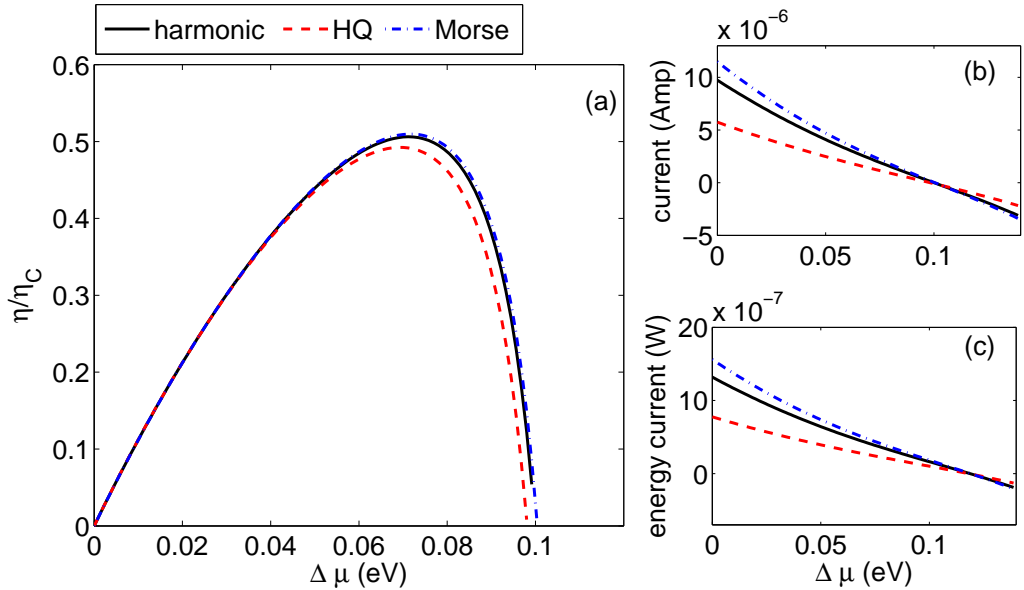}  
\caption{
(a) Thermoelectric efficiency $\eta/\eta_C$ far from equilibrium with
$\eta_C=1-T_C/T_H$
(b) Charge current $\langle I_p\rangle $ and (c) energy current $\langle I_e\rangle$
for the harmonic (full), HQ (dashed), and Morse (dashed-dotted) potentials with
$\omega_0$ = 0.05, $\epsilon_0$ = 0.15, $g$ = 0.01, $\Gamma$ = 0.1, $\Gamma_{ph} = 0$, in units of
eV, and $T_L$ = 300 K, $T_R$ = 800 K.
}
\label{NLR1}
\end{figure*}

We explore the thermoelectric efficiency beyond linear response in
Fig. \ref{NLR1} where we display the charge and energy currents across the junction,
 along with the energy conversion efficiency, as a function of applied bias
for $\Delta T=500$ K.
The Morse oscillator supports higher currents than the harmonic-oscillator and the HQ models (up to a factor of two),
but the thermoelectric efficiency only mildly deviates between the three cases.
Since many levels contribute to the currents at this high temperature-high bias limit, we cannot put forward a simple argument
justifying this correspondence. We know however, from analytical considerations,
 that the harmonic oscillator case and the
two-state mode build up an {\it identical} thermoelectric energy conversion efficiency \cite{BijayPRB15}.
Additional work is required to clarify on this correspondence in highly-biased, high-$T$, genuinely anharmonic models.

\section{Conclusion}
\label{Summary}

We studied the electrical transport characteristics and thermoelectric efficiency of
a phonon-assisted donor-acceptor junction, focusing on the role of the vibrational potential on transport behavior.
We demonstrated that the inelastic current can reveal signatures of molecular anharmonicity, e.g., showing new peaks
in the differential conductance, the result of compromised harmonic selection rules. In contrast,
properties that depend on the ratio of the (inelastic) charge and energy currents, such as the
thermopower and the thermoelectric efficiency, only mildly reveal the underlying molecular anharmonicity.
The thermopower and the thermoelectric efficiency could be tuned by modifying the electronic parameters,
$\Gamma$ and $\epsilon_{d,a}$ \cite{BijayDABeil}. However, the nature of the nuclear motion only lightly influences these
quantities. We emphasize though that our calculations do not include the process of phononic thermal conduction across the junction,
a factor that can significantly affect the overall efficiency \cite{Fabian2}. 
Other contributions here include the organization of working expressions for the inelastic current (\ref{eq:curr1})-(\ref{eq:curr2}),
and the derivation of a scaling law for the charge current at high bias.

Our calculations were performed with a quantum master equation which is perturbative in
the electron-vibration coupling but exact to all order in the metal-molecule hybridization \cite{SiminePCCP}.
This should be contrasted with other QME methods which are developed based on the exact treatment of
electron-vibration interaction while including the metal-molecule coupling as perturbative parameter \cite{Jens,Mitra,Esposito-QME,wegeK,Peskin}.
QME methods can handle vibrational anharmonicities in an exact manner
unlike the non-equilibrium Green's function (NEGF) technique, a complementary perturbative treatment
\cite{NEGF-Rammer,NEGF-rev-Bijay}.
While we do not have a benchmark for our analysis here---with anharmonic potentials---
in Ref. \cite{Bijay16} we showed that our QME can be exercised in a compatible manner with an NEGF method, in a
junction with a harmonic vibrational mode.

Our method is flexible: It can handle for example nonlinear interactions in the form
 $\hat S = e^{-\alpha \hat x}$, as examined in Ref. \cite{VonOppen},
since matrix elements $S_{m,n}$ can be reached numerically. We can also use our method and simulate transport junctions with several-prominent vibrations.
Finally, the QME as described here can be used to examine a range of transport problems, by turning on/off different reservoirs.
Besides the analysis of inelastic electronic conduction with anharmonic modes, one can use this method
and study the operation of phonon-thermoelectric transistors \cite{Jiang1} and
phononic thermal junctions with harmonic and anharmonic local modes, to demonstrate nonlinear function such as thermal rectification and negative differential thermal conductance \cite{SB}.

\section*{Acknowledgments}
This work was funded by an NSERC Discovery Grant, the Canada Research Chair program,
and the CQIQC at the University of Toronto.


\renewcommand{\theequation}{A\arabic{equation}}
\setcounter{equation}{0}  
\section*{Appendix A: Derivation of Eqs. (\ref{eq:curr2}) and (\ref{eq:curre2}) } 

\begin{widetext}

To derive Eq.  (\ref{eq:curr2}) for the charge current,
our starting point is equation (\ref{eq:curr1}) with the
rate constants (\ref{eq:kFe}),
\bea
k_{n \to m}^{\nu\to \bar \nu}=
| S_{m,n}|^2 \int_{-\infty}^{\infty} d\tau e^{i (E_n-E_m) \tau}  \langle \hat B_{\nu}(\tau)\, \hat B_{\bar \nu}(0) \rangle.
\nonumber\\
\label{eq:AkF}
\eea
Here  $\hat B_L\equiv g\sum_{r,l} \gamma_{l}^*\gamma_r \hat a_l^{\dagger}\hat a_r$, and similarly
$\hat B_R=g\sum_{r,l} \gamma_{r}^*\gamma_l \hat a_r^{\dagger}\hat a_l$,  $\hat B_R=\hat B_L^{\dagger}$.
Averages are performed with respect to the grand-canonical state in the $L$ and $R$ leads.
We can now organize the following expression,
%
\bea
&&\sum_{m,n}p_n^{ss} k_{n\to m}^{R\to L}
= \int_{-\infty}^{\infty}d\tau \sum_{m,n}p_n^{ss} |S_{m,n}|^2 e^{-iE_{nm}\tau} \langle \hat B_R(0)\hat B_L(\tau)\rangle
\nonumber\\
&=&\int_{-\infty}^{\infty} d\tau \langle \hat S(0)\hat S(\tau)\rangle_{ss}\langle \hat B_R(0)\hat B_L(\tau)\rangle
=
\int_{-\infty}^{\infty} d\tau C_S(\tau)C_{RL}(\tau)
=
\frac{1}{2\pi} \int d\omega C_S(\omega) C_{RL}(-\omega),
\label{eq:A1}
\eea
%
with
\bea
C_S(\tau)=
\sum_n p_n^{ss} \langle n|  \hat S(0)\hat S(\tau)|n\rangle.
\eea
In frequency domain,
\bea
C_S(\omega)=2\pi \sum_{n,m} p_n^{ss} |\langle m| \hat S|n\rangle|^2 \delta(\omega+E_{mn}).
\label{eq:CSw}
\eea
The bath correlation functions are
 $C_{\nu,\bar \nu}(\tau)\equiv \langle \hat B_{\nu}(0)\hat B_{\bar \nu}(\tau)\rangle$,
 $\nu,\bar \nu=L,R$,
or explicitly,
\bea
C_{RL}(\tau) &=& \langle B_{R}(0) B_{L}(\tau)\rangle =
g^2 \sum_{l,r}|\gamma_{l}|^2|\gamma_r|^2 f_R(\epsilon_r) [1-f_L(\epsilon_l)] e^{i(\epsilon_l-\epsilon_r)\tau},
\nonumber\\
C_{RL}(\omega)
&=& \frac{1}{2\pi} \int_{-\infty}^{\infty} d\epsilon f_R(\epsilon)\left[1-f_L(\epsilon-\omega) \right]J_R(\epsilon)J_L(\epsilon-\omega),
\eea
with  $J_{L,R}(\omega)$ given in Eq. (\ref{eq:spec}).
Similarly, the second expression in Eq. (\ref{eq:curr1}) organizes to
\bea
\sum_{m,n}p_n^{ss} k_{n\to m}^{L\to R}
=\int_{-\infty}^{\infty} d\tau C_S(\tau)C_{LR}(\tau)
=
\frac{1}{2\pi} \int_{-\infty}^{\infty} d\omega C_S(\omega) C_{LR}(-\omega).
\nonumber\\
\label{eq:A2}
\eea
Combining Eq. (\ref{eq:A1}) with (\ref{eq:A2}), we arrive at Eq. (\ref{eq:curr2}) for the charge current
\bea
\langle I_p \rangle = \sum_{n,m}p_n^{ss}\left[  k_{n\to m}^{R\to L} - k_{n\to m}^{L\to R}\right]
= \frac{1}{2\pi} \int_{-\infty}^{\infty} d\omega C_S(\omega)[ C_{RL}(-\omega)-  C_{LR}(-\omega)].
\eea
By following similar steps, we derive next Eq. (\ref{eq:curre2}) for the energy current.
We begin from Eq. (\ref{eq:Ecurrent-QME}),
\bea
\langle I_e \rangle =  \sum_{m,n} p_n^{ss} \frac{\partial k_{n \to m}^{\lambda}}{\partial (i\lambda_e) } \Big|_{\lambda=0},
\label{eq:AEcurrent-QME}
\eea
with the energy relaxation/excitation rate constants
\bea
&&\frac{\partial [k_{n\to m}^{\lambda}]^{L\to R}}{\partial {(i\lambda_e)}}{\Bigg|}_{\lambda=0}
= | S_{m,n}|^2\int_{-\infty}^{\infty} \frac{d\epsilon}{2\pi}  [-\epsilon - E_{nm}] f_L(\epsilon) (1-f_R(\epsilon+E_{nm})) J_L(\epsilon) J_R(\epsilon+E_{nm}),
\nonumber \\
&&\frac{\partial [k_{n\to m}^{\lambda}]^{R \to L}}{\partial {(i\lambda_e)}}{\Bigg|}_{\lambda=0}=
| S_{m,n}|^2
\int_{-\infty}^{\infty} \frac{d\epsilon}{2\pi}\epsilon f_R(\epsilon) (1-f_L(\epsilon+E_{nm})) J_R(\epsilon) J_L(\epsilon+E_{nm}).
\eea
In analogy with Eq. (\ref{eq:AkF}), we introduce the following definitions,
\bea
\dot k_{n \to m}^{R\to L}&=&
| S_{m,n}|^2 \int_{-\infty}^{\infty} d\tau e^{-i (E_n-E_m) \tau}  \langle \hat B_{R}(0)\, \dot{\hat B}_{L}(\tau) \rangle,
\nonumber\\
&=&
i| S_{m,n}|^2
\int_{-\infty}^{\infty} \frac{1}{2\pi}d\epsilon (E_{nm} +\epsilon) f_R(\epsilon)[1-f_L(\epsilon+E_{nm}) J_R(\epsilon) J_L(\epsilon+E_{nm})
\nonumber\\
\dot k_{n \to m}^{L\to R}&=&
| S_{m,n}|^2 \int_{-\infty}^{\infty} d\tau e^{-i (E_n-E_m) \tau}  \langle \dot{\hat B}_{L}(0) \hat B_{R}(\tau) \rangle,
\nonumber\\
&=&
i| S_{m,n}|^2
\int_{-\infty}^{\infty}\frac{1}{2\pi}d\epsilon \epsilon f_L(\epsilon)[1-f_R(\epsilon+E_{nm}) J_L(\epsilon) J_R(\epsilon+E_{nm}).
\label{eq:AkFdt}
\eea
The time derivative corresponds to
$\dot{\hat B}_L(t)=i[\hat H_L(t),\hat B_L(t)] = g\sum_{l,r}\epsilon_l\gamma_{l}^*\gamma_r \hat a_l^{\dagger}(t)\hat a_r(t)$
with the time evolution given in the interaction representation.
Using Eq. (\ref{eq:AEcurrent-QME}), we construct the energy current
\bea
\langle I_e \rangle &=& i \sum_{n,m} p_{n}^{ss} \left[    \dot k_{n\to m}^{R\to L}  - \dot k_{n\to m}^{L\to R}\right]
\nonumber\\
&=&
i\int_{-\infty}^{\infty} d\tau C_{S}(\tau) \left[\dot C_{RL}(\tau) -\dot C_{LR}(\tau)\right]
= \frac{1}{2\pi} \int_{-\infty}^{\infty} d\omega \omega C_S(-\omega)\left[C_{RL}(\omega)-C_{LR}(\omega)\right].
\eea

\end{widetext}


\renewcommand{\theequation}{B\arabic{equation}}
\setcounter{equation}{0}  
\section*{Appendix B: Elastic contribution to the charge current}

%
\begin{figure*}[htbp]
\hspace{-21mm}
\includegraphics[width=14cm]{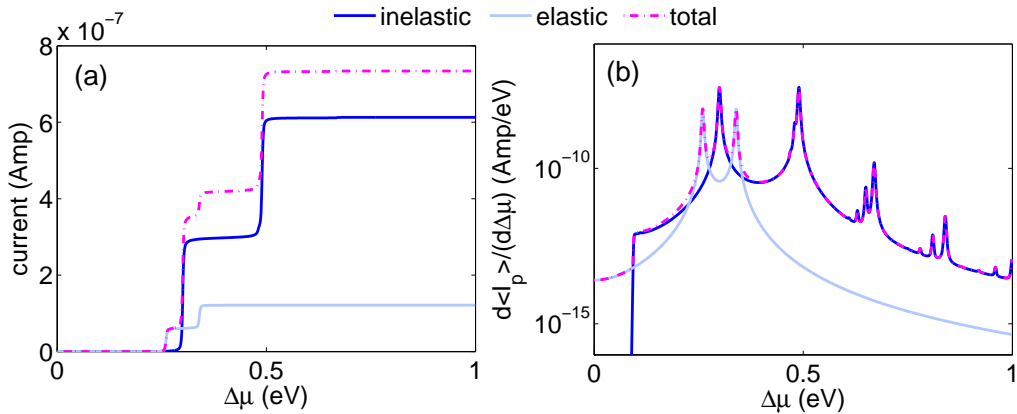}
\caption{(a) Elastic (light), inelastic (full) and total (dashed-dotted) charge currents
and (b) their first derivative with respect to bias.
We used the Morse potential with same parameters as in Fig. \ref{IVAH1},
including a tunneling energy $t=0.02$ eV.
}
\label{AppB}
\end{figure*}

The current from electrons that transverse the system elastically and coherently can be included
using the Landauer formalism. 
The steady state charge current, defined as positive from right to left, is expressed as,
\bea
I_{elastic} = \frac{e}{2 \pi \hbar} \int_{-\infty}^{\infty} d\epsilon \, \mathcal{T}(\epsilon) [f_R(\epsilon)-f_L(\epsilon) ],
\label{landauer_formula}
\eea
with $f_{\nu}(\epsilon)$ as the Fermi-Dirac function.
The transmission probability  $\mathcal{T}(\epsilon)$ can be obtained from the Green's function formalism using,
\bea
\mathcal{T}(\epsilon) = \textrm{Tr}[\hat{\mathcal{G}}^{r}(\epsilon) \hat \Gamma_L \hat{\mathcal{G}}^{a} (\epsilon) \hat \Gamma_R],
\label{trace_trans}
\eea
where $\hat{\mathcal{G}}^{r}(\epsilon)$ is the retarded Green's function,
\bea
\hat{\mathcal{G}}^{r}(\epsilon) = [\hat{I} \epsilon - \hat{H}_M + i \hat{\Gamma}/2]^{-1},
\label{greenz}
\eea
and $\hat{\mathcal{G}}^{a}(\epsilon) = [\hat{\mathcal{G}}^{r}(\epsilon)]^{\dagger}$. $\hat{\Gamma}_{L,R}$ are hybridization matrices for left and right leads,
\bea
\hat{\Gamma}_{L} =
\begin{bmatrix}
    \Gamma_L & 0 \\
    0        & 0 \\
\end{bmatrix},
\hat{\Gamma}_{R} =
\begin{bmatrix}
    0 & 0        \\
    0 & \Gamma_R \\
\end{bmatrix}
\eea
with $\Gamma_{L,R}$ as the lead-molecule hybridization for left or right leads, taken to be equal throughout the text as $\Gamma$.
We define the terms in Eq. (\ref{greenz}) as follows. $\hat{I}$ is the identity matrix, $\hat{\Gamma}$ is a sum of the two hybridization matrices ($\hat{\Gamma} = \hat{\Gamma}_L + \hat{\Gamma}_R)$ and $\hat{H}_M$ is the molecular Hamiltonian,
\bea
\hat{H}_M =
\begin{bmatrix}
    \epsilon_d & t        \\
    t          & \epsilon_a \\
\end{bmatrix}
\eea
Recall that $\epsilon_0=\epsilon_{d,a}$ are the D and A energy levels, $t$ is the tunneling energy between D and A.
 Using Eq. (\ref{trace_trans}) and the above definitions we find the transmission function
\bea
\mathcal{T}(\epsilon) = \frac{t^2 \Gamma_L \Gamma_R}{| (\epsilon-\epsilon_0+i \Gamma_L/2)(\epsilon-\epsilon_0+i \Gamma_R/2)-t^2 |^2}.
\eea
Employing Eq. (\ref{landauer_formula}), we calculate the elastic current with a range of applied voltages to obtain Fig. \ref{AppB}(a). We make the non-crossing approximation and write down the total current
as the sum of the elastic and inelastic currents,
%
%
which gives the dashed-dotted curve in Fig. \ref{AppB}(a).
The inelastic curve (dark) is identical to the one included in Fig. \ref{IVAH1}(a).
The differential conductance is shown in Fig. \ref{AppB} (b). We observe two new peaks in the first derivative of the total current corresponding to the added elastic processes which occur at biases that satisfy the resonance conditions,
\bea
&& \Delta \mu \approx 2(\epsilon_0 + t) = 0.34 \textrm{ eV} \nonumber \\
&& \Delta \mu \approx 2(\epsilon_0 - t) = 0.26 \textrm{ eV}.
\eea
Since these are elastic processes, electrons do not exchange there energy with the molecular vibration, unlike the peak at 0.3 eV coming up from the inelastic contribution.


\begin{thebibliography}{99}
\bibitem{commGaph}
The spectral function $\Gamma_{ph}(\omega)$ in general depends on frequency, modelled e.g. with an Ohmic function.
Here, to focus on the role of the vibrational spectrum on transport, we assumed it to be a constant.
Since $\Gamma_{ph}$ affects our results only at high bias when mode-heating is significant, this choice only mildly influences our
 simulations.
\end{thebibliography}
\end{document}